\documentclass[aps,10pt,amsmath,amssymb]{revtex4}
\usepackage{amsfonts}
\usepackage{amssymb}
\usepackage{graphicx}
\usepackage{subfigure}
\usepackage{color}
\usepackage[normalem]{ulem}
\usepackage{bbold}
\usepackage{placeins}
\usepackage{soul,xcolor}
\usepackage{epstopdf}
\usepackage{tabu}
\usepackage{array}
\usepackage{upgreek}
\usepackage{multirow}
\usepackage{braket}
\usepackage[utf8]{inputenc}
\usepackage[colorlinks = truelinkcolor = blue, urlcolor  = blue, citecolor = blue, anchorcolor = blue]{hyperref}
\hypersetup{
	colorlinks   = true, 
	urlcolor     = blue, 
	linkcolor    = blue, 
	citecolor   = blue 
}

\begin{document}

\title{Entropic Leggett-Garg inequality in neutrinos and $B (K)$ meson systems}

\author{Javid Ahmad Naikoo}
\email{naikoo.1@iitj.ac.in}
\affiliation{Indian Institute of Technology Jodhpur, Jodhpur 342011, India}

\author{Subhashish Banerjee}
\email{subhashish@iitj.ac.in}
\affiliation{Indian Institute of Technology Jodhpur, Jodhpur 342011, India}

\begin{abstract}
Entropic Leggett-Garg inequality is studied in systems like neutrinos in the  context of two and three flavor neutrino oscillations and in neutral $B_d$, $B_s$ and $K$ mesons. The neutrino dynamics is described with the matter effect taken into consideration. For the decohering $B/K$ meson systems, the effect of decoherence and CP violation have also been taken into account, using the techniques of open quantum systems.  Enhancement in the violation with increase in the number of measurements has been found, in consistency with findings  in spin-$s$ systems. The effect of decoherence is found to bring the deficit parameter $\mathcal{D}^{[n]}$ closer to its classical value zero, as expected. The violation of entropic Leggett-Garg inequality lasts for a much longer time in $K$ meson system than in $B_d$ and $B_s$ systems.
\end{abstract}
\maketitle

\section{INTRODUCTION}\label{intro}
There is no sharp boundary between the classical and the quantum worlds. However, physicists have developed some important notions which can shed light on the distinction of the two domains. The most profound among these notions being the uncertainty principle \cite{Heisenberg}. The violation of Bell inequality \cite{Bell:1964kc, PhysRevLett.23.880} is another prominent example which reveals the nonclassical nature  of correlations between spatially separated quantum systems \cite{RevModPhys.65.803}. Aspect's experiment \cite{PhysRevLett.49.91}  verified for the first time the CHSH form of the Bell inequality by using pairs of spatially separated polarization-entangled photons. Since then, Bell theorem has been successfully verified in many experiments  \cite{PhysRevA.57.3229, PhysRevLett.81.3563, PhysRevLett.81.5039, pan2000experimental, rowe2001experimental, PhysRevLett.100.220404, kofler2013bell, Handsteiner:2016ulx, PhysRevLett.119.010402}.\par
Quantum correlations could be spatial or temporal. Among spatial quantum correlations, much attention has been devoted to entanglement \cite{horodecki2009quantum}: entangled states are  non classical and sometimes  display even stronger correlations such as steering \cite{cavalcanti2016quantum} and non-locality \cite{brunner2014bell}. However, even unentangled states, may not have a classical description.  Quantum discord \cite{ollivier2001quantum, henderson2001classical, adhikari2012operational} is another important spatial quantum correlation and captures the fact that local measurements on parts of a composite system induce an overall disturbance in the state. Details of various facets of spatial quantum correlations studied on different physical systems can be seen in, for example, \cite{aspect1981experimental, tittel1998experimental, tittel1998violation, PhysRevLett.81.5039,BANERJEE2010816, chakrabarty2010study,lanyon2013experimental, kessel2000quantum, laflamme2001nmr, blasone2009entanglement, alok2016quantum, banerjee2015quantum, banerjee2016quantum, khushQC}.\par
Leggett-Garg inequalities (LGIs), considered to be the temporal analogue of  Bell inequalities, were constructed to understand the extrapolation of quantum theory to  macroscopic systems, by assuming \textit{macro-realism} and \textit{noninvasive measurability} \cite{leggett1985quantum}. \textit{Macro-realism} implies that a system with two or more distinct states available to it will be, at any time, in one of these states.  \textit{Noninvasive measurability} means that it is possible to determine this state without disturbing the future dynamics of the system.  LGIs have been studied in various theoretical works \cite{barbieri2009multiple, avis2010leggett, lambert2010distinguishing, lambert2011macrorealism, montina2012dynamics, emary2013leggett, PhysRevA.87.052115,emarydeco} including, in recent times, neutrino oscillations \cite{ Naikoo:2017fos, Fu:2017hky} and neutral mesons \cite{JavidLGImeson}  and verified in a number of experiments  \cite{palacios2010experimental, groen2013partial, goggin2011violation, xu2011experimental, dressel2011experimental, suzuki2012violation, athalye2011investigation, souza2011scattering, katiyar2013violation}. The Leggett-Garg string for an $n$ measurement scenario is given as
\begin{equation}\label{LG-n}
K_n =\sum\limits_{i=1}^{n} C_{i,i+1} - C_{i,n}.
\end{equation}
Here, $C_{ij} = \langle \hat{A}(t_i) \hat{A}(t_j) \rangle$ is the two time correlation function for the operator $\hat{A}$. The assumptions of macrorealism and non-invasive measurability impose the following restrictions on $K_n$:
\begin{align}
-n &\le K_n \le n-2 \qquad \qquad{\rm for}~  n\ge 3, n ~~odd; \nonumber\\
-(n-2) &\le K_n \le n-2 \qquad \qquad{\rm for}~ n\ge 4, n~~even.
\end{align}
The entropic version of Bell inequality was formulated in \cite{braunstein1988information}. Entropic formulations derive their utility from their ability to deal with any finite number of outcomes, allowing, in principle, to go beyond the standard dichotomic choice of observable \cite{chaves2012entropic, rastegin2014formulation}. Entropic version of the temporal counterpart of Bell inequality was developed in \cite{morikoshi2006information}. This was followed by  an application of the  entropic version of Leggett-Garg inequality to a spin-$s$ system in \cite{devi2013macrorealism}.\par
In this work, we will analyze the entropic Leggett-Garg inequality for the neutrino system in the context of neutrino oscillations and for the decohering $K$ and $B$ mesons by using the formalism of open quantum systems. In Sec. (\ref{LGI-section}), we provide a brief account of the entropic formulation of Leggett-Garg inequality. Section (\ref{Neutrino-Osc}) is devoted to a discussion of neutrino oscillations and the entropic Leggett-Garg inequality for the neutrino system.  The dynamics of  $K$ and $B$ mesons using the open systems approach is spelled out in Sec. (\ref{Meson-Dynamics}) wherein we also discuss the construction of entropic Leggett-Garg inequality for the meson system.  Section (\ref{results})  is devoted to a discussion of  the results obtained. We  conclude  in Sec. (\ref{conclusion}).

\section{Entropic Leggett-Garg inequality}\label{LGI-section}
We now provide a brief review of some rudiments of information theory used in the development of the entropic Leggett-Garg inequality. We begin by considering the observable $A$ which can take discrete values denoted by $a_i$ at time $t_i$, that is, $A(t_i) = a_i$. We define the joint probability  of the measurement of $A$ at times $t_i$ and $t_j$ giving results $a_i$ and $a_j$, respectively, as $P(a_i, a_j)$. According to Bayes's theorem the joint probability is related to the conditional probability as,
\begin{equation}\label{Bayes}
P(a_i, a_j)  = P(a_j|a_i) P(a_i) = P(a_i|a_j) P(a_j).
\end{equation}
Here, $P(a_j|a_i)$ is the conditional probability of obtaining the outcome $a_j$ at time $t_j$, given that $a_i$ was obtained at time $t_i$.\par
A classical theory can assign well defined values to all observables of the system with no reference to the measurement process. This assumption lies at the heart of Bell and Leggett-Garg inequalities, leading to bounds which may  not be respected by the non-classical systems. In other words, this assumption  demands a joint probability distribution,\\ $P(a_i, a_j)$, yielding  information about the marginals of individual observations at time $t_i$. The assumption of non-invasive measurability implies that the measurement made on a system at any time does not disturb its future dynamics and hence any measurement made at a later time $t_j$ where $t_j>t_i$. The mathematical statement would be that the joint probabilities be expressed as a convex combination of the product of probabilities $P(a_i|\lambda)$, averaged over a hidden variable probability distribution $\rho(\lambda)$ \cite{PhysRevA.87.052115, kofler2008conditions, PhysRevLett.48.291}:
\begin{equation}
P(a_1, a_2, \dots, a_n) = \sum\limits_{\lambda} \rho(\lambda) P(a_1|\lambda) P(a_2|\lambda) \cdots P(a_n|\lambda),
\end{equation}
such that the following properties are satisfied
\begin{equation}
0 \le \rho(\lambda) \le 1, \quad \sum\limits_{\lambda} \rho(\lambda) = 1;  \quad 0\le P(a_i|\lambda)\le 1, \quad \sum\limits_{\lambda} P(a_i|\lambda) =1.
\end{equation}
One can use the conditional probability given by Eq. (\ref{Bayes}) to define the conditional entropy as
\begin{equation}\label{cond-entropy}
H[A(t_j)|A(t_i)] = -\sum\limits_{a_i, a_j} P(a_j|a_i) \log_2 P(a_j|a_i).
\end{equation}
Now using chain rule and the fact that conditioning reduces entropy \cite{cover2012elements}, one obtains \cite{morikoshi2006information}
\begin{equation}
H[A(t_{N-1}),...,A(t_0)]  \le H[A(t_{N-1})|A(t_{N-2})] + ... + H[A(t_1)|A(t_{0})] + H[A(t_0)].
\end{equation}
This temporal entropic inequality was used in \cite{morikoshi2006information} to study the role of quantum coherence in Grover's algorithm. Using the relation $H[A(t_i), A(t_{i+j})] = H[A(t_{i+j})|A(t_i)] + H[A(t_i)]$, one can derive the temporal analogues of the spatial entropic Bell inequalities 
\begin{equation}\label{Gen-Ineq}
\sum\limits_{k=1}^{N-1} H[A(t_k)|A(t_{k-1})] - H[A(t_{N-1})|A(t_0)] \ge 0.
\end{equation}
These are the entropic Leggett-Garg inequalities \cite{devi2013macrorealism}. Here, $N$ denotes the number of measurements, inclusive of the preparation; the case of $N=3$ was experimentally tested in  \cite{katiyar2013violation}.

\section{Neutrino oscillation} \label{Neutrino-Osc}
Consider an arbitrary number of $n$ orthogonal flavor eigenstates $\ket{\nu_\alpha}$ with $\langle \nu_\alpha | \nu_\beta \rangle = \delta_{\alpha \beta}$. These flavor states are connected to $n$ mass eigenstates by a unitary operator $U$ \cite{bilenky1999phenomenology, gonzalez2008phenomenology}: 
\begin{equation}
\ket{\nu_\alpha} =  \sum\limits_{k} U_{\alpha k} \ket{\nu_k} \qquad \ket{\nu_k} = \sum\limits_{\alpha} U^{*}_{\alpha k} \ket{\nu_\alpha}, \label{nu-alpha}
\end{equation}
such that $U^\dagger U=U U^\dagger=\mathbb{1} $, $\sum\limits_{k} U_{\alpha k}^* U_{\beta k}=\delta_{\alpha \beta}$ and $\sum\limits_{\alpha} U_{\alpha i}^* U_{\alpha k}=\delta_{i k}$. The mass eigenstates are stationary states and have the following time dependence:
\begin{equation}
\ket{\nu_k (t)} = e^{-i E_k t} \ket{\nu_k(0)}. \label{nu-k}
\end{equation}
From Eqs. (\ref{nu-alpha}) and (\ref{nu-k}) we conclude
\begin{align}
\ket{\nu_\alpha (t)} &= \sum\limits_{k} U_{\alpha k} e^{-i E_k t} \ket{\nu_k(0)},\nonumber \\ 
&= \sum\limits_{k} \sum\limits_{\beta} U_{\alpha k} e^{-i E_k t} (U_{\beta k})^* \ket{\nu_\beta(0)}.
\end{align}
An arbitrary neutrino state $\psi \in \mathcal{H}$ can be expanded in both  the flavor and mass basis as \cite{ohlsson2000three}:
\begin{align}
\ket{\psi} &\equiv \sum\limits_{\alpha=e, \mu, \tau} \psi_\alpha \ket{\nu_\alpha} = \sum\limits_{\alpha=e, \mu, \tau} \psi_\alpha \bigg(\sum_{k=1}^{3} U^*_{\alpha k} \ket{\nu_k}\bigg),\nonumber \\
&= \sum\limits_{k=1}^{3} \bigg(\sum\limits_{\alpha=e, \mu, \tau} \psi_\alpha U^*_{\alpha k} \bigg) \ket{\nu_k}\equiv \sum\limits_{k=1}^{3} \psi_k\ket{\nu_k}.
\end{align}
Here $\psi_\alpha$ and $\psi_k$ are the components of the wavefunction $\ket{\psi}$ in the flavor basis and the mass eigenbasis, respectively, with the following relation:
\begin{align}
\psi_\alpha &= \sum\limits_{\alpha=e, \mu, \tau} U_{\alpha k} \psi_k \qquad k=1, 2, 3.\\
\psi_f &= U \psi_m,
\end{align}
with the subscripts $f$ and $m$ denoting the flavor and the mass state basis, respectively. 
For three flavor scenario ($n=3$), a convenient parametrization for $U(\theta_{12},\theta_{23},\theta_{32},\delta)$ is given by  \cite{chau1984comments, giunti2007fundamentals} \\ \\
\begin{widetext}
\begin{equation}
U(\theta_{12},\theta_{23},\theta_{32},\delta) = 
\begin{pmatrix}
c_{12} c_{13} & s_{12} c_{13} & s_{23} e^{-i \delta} \\ - s_{12}c_{23} - c_{12} s_{23}s_{13} e^{i\delta} & c_{12}c_{23}-s_{12}s_{23}s_{13} e^{i\delta} & s_{23}c_{13} \\ s_{13}s_{23} - c_{12}c_{23}s_{13} e^{i\delta} & -c_{12}s_{23}-s_{12}c_{23}s_{13} e^{i\delta} & c_{23}c_{13}\end{pmatrix},
\end{equation}
\end{widetext}
where $c_{ij} = \cos\theta_{ij}$, $s_{ij} = \sin\theta_{ij}$,  $\theta_{ij}$ are the mixing angles and $\delta$ the $CP$ violating phase.\\
We are usually interested in the unitary transformation $U_f$ that would take the state $\psi(0)$ at time $t=0$ to $\psi(t)$ at a later time $t$, that is, $\psi(t) = U_f \psi(0)$, where $U_f = e^{-i H_f t} = U e^{-i H_m t} U^{-1} = e^{-iH_f t}$ \cite{ohlsson2000three}, with $H_f = U H_m U^{-1}$  holds for the case of neutrinos traveling in vacuum and the flavor evolution matrix $U_f$ can be expressed in a compact form. A detailed account of dealing with the neutrinos propagating through a matter density is given in \cite{barger1980matter, kim1987adiabatic, zaglauer1988mixing}. An entirely different approach was developed for treating neutrino oscillations in presence of matter effect in \cite{ohlsson2000three, Ohlsson:1999um, Ohlsson:2001et}.\par

$  Entropic~Leggett-Garg~Inequality~ for~ Neutrinos:$\par

Let us assume that we have prepared an ensemble of neutrinos all existing in a \textit{fixed} flavor state, say $\nu_\alpha$  ($\alpha = e, \mu, \tau$), at time $t_i$. We choose the projector $\Pi = | \nu_\beta \rangle \langle \nu_\beta|$, which projects a particular flavor state $\ket{\nu_\beta}$ ($\beta = e, \mu, \tau$). In Heisenberg picture, $\Pi(t) = U_f^{\dagger}(t) \Pi U_f(t)$. For brevity, let us use the notation $\alpha_j$ to denote the flavor state $\ket{\nu_\alpha}$ at time $t_j$. The conditional probability of obtaining outcome $\alpha_{j+1}$ at time $t_{j+1}$ given that $\alpha_j$ was obtained at time $t_j$ is given by		
\begin{align}
P(\alpha_{j+1},t_{j+1} | \alpha_j,t_j) &= Tr[\rho^\prime \Pi_{\alpha_{j+1}}(t_{j+1})],\nonumber  \\
&= Tr\Bigg[ \frac{ \Pi_{\alpha_{j}}(t_{j}) \rho(0) \Pi_{\alpha_{j}}(t_{j}) }{Tr[\rho(0) \Pi_{\alpha_j}(t_j)]} \Pi_{\alpha_{j+1}}(t_{j+1}) \Bigg],\nonumber \\
&= Tr\Bigg[ \frac{ \Pi_{\alpha_{j}}(t_{j}) \rho(0) \Pi_{\alpha_{j}}(t_{j}) }{P_{\alpha_j}(t_j)} \Pi_{\alpha_{j+1}}(t_{j+1}) \Bigg].
\end{align} 
Here $ \rho^\prime$ is the state after the projective measurement made at time $t_j$ and is given by $\frac{ \Pi_{\alpha_{j}}(t_{j}) \rho(0) \Pi_{\alpha_{j}}(t_{j}) }{Tr[\rho \Pi_{\alpha_j}(t_j)]}$. The joint probability, therefore becomes
\begin{widetext}
\begin{align}
P(\alpha_j,\alpha_{j+1})  &=  Tr[\Pi_{\alpha_j}(t_j) \rho(0) \Pi_{\alpha_j}(t_j) \Pi_{\alpha_{j+1}}(t_{j+1}) ],\nonumber \\
&=  Tr[ U_f^\dagger(t_j) \ket{\alpha_j}\bra{\alpha_j}  U_f(t_j)  \rho(0) U_f^\dagger(t_j) \ket{\alpha_j}  \bra{\alpha_j}  U_f(t_j) U_f^\dagger (t_{j+1} )  \ket{\alpha_{j+1}}\bra{\alpha_{j+1}} U_f(t_{j+1}) ],\nonumber \\
&=  Tr[ \ket{\alpha_j}\bra{\alpha_j}  U_f(t_j)  \rho(0) U_f^\dagger(t_j) \ket{\alpha_j}  \bra{\alpha_j}  U_f(t_j) U_f^\dagger (t_{j+1} )   \ket{\alpha_{j+1}}\bra{\alpha_{j+1}} U_f(t_{j+1})   U_f^\dagger(t_j) ],\nonumber \\
&= \langle \alpha_j | \rho(t_j) | \alpha_j \rangle ~| \langle \alpha_{j+1} | U_f(t_{j+1}) U_f^{\dagger}(t_j) | \alpha_j \rangle |^2.
\end{align} 
\end{widetext}
This joint probability can be used to compute the mean conditional information entropy
\small
\begin{equation}
H(Q_{k+1}|Q_k) = -\sum\limits_{\alpha_k, \alpha_{k+1}} P(\alpha_{k+1},\alpha_k) \log_2 \bigg( \frac{P(\alpha_{k+1},\alpha_k)}{P(\alpha_k)} \bigg).
\end{equation}
\normalsize
 Here $\alpha_k$ is a particular realization of the random variable $Q_k$. For a neutrino born in flavor state $|\nu_\alpha \rangle$ at time $t_0$, we have  $\rho(t_0) = | \nu_\alpha \rangle \langle \nu_\alpha |$, and 	

\begin{align}
H(Q_{t_1}|Q_{t_0}) &=	H[\nu(t_1)|\nu(t_0) = \nu_{\alpha}] \nonumber \\
&=- \mathcal{P}_{\alpha \alpha}(t_1 -t_0)\log_2\mathcal{P}_{\alpha \alpha}(t_1 -t_0)  - \sum\limits_{\beta \neq \alpha} \big[ \mathcal{P}_{\alpha \beta}(t_1 -t_0)\log_2 \mathcal{P}_{\alpha \beta}(t_1 -t_0)  \big].
\end{align}
Here $\mathcal{P}_{\alpha \alpha}(t_1 -t_0)$ and $\mathcal{P}_{\alpha \beta}(t_1 -t_0)$ are for the survival  and  transition  probability, respectively.\par
Given an ensemble of identically prepared neutrinos at time $t_0$, and considering the preparation step as the first measurement,  we can perform a series of measurements, for (say) $N=3$, such that on the first set of runs, the measurement is performed at time $t_1$;  only at $t_1$ and $t_2$ on the second set of runs and at $t_2$ on the third run $(t_2>t_1>t_0)$. For measurements carried out at equal time intervals $\Delta t = t_{i+1} - t_i$, $i=1,2,\dots, n$, the survival and oscillation probabilities depend only on the time difference $\Delta t$. We define a dimensionless parameter  $\phi$, which is related to $\Delta t$ as $\phi = (\Delta_{21} \Delta t)/(2\hbar E)$, where $\Delta_{21} = m_2^2 - m_1^2$ is the mass squared difference and E is the energy of the neutrino.\par 
As an example, the mean conditional information, when the initial state at time $t_0$ is chosen to be $\ket{\nu_e}$, as a function of   $\phi$ has the following form:
\begin{align}
H[\nu(t_1)|\nu(t_0)=\nu_{e}](\phi) 	&= - \mathcal{P}_{e e}(\phi)\log_2\mathcal{P}_{e e}(\phi)  - \big[ \mathcal{P}_{e \mu}(\phi)\log_2 \mathcal{P}_{e \mu}(\phi)  +  \mathcal{P}_{e \tau}(\phi)\log_2\mathcal{P}_{e \tau}(\phi) \big]  \label{H10-eflavor}.
\end{align}
Similarly, one can find the expressions for $H(Q_2|Q_1)$ and $H(Q_2|Q_0)$. It turns out that the actual form of $H(Q_2|Q_1)$ involves  probabilities which cannot be measured with the present day neutrino experimental facilities. One can overcome this difficulty by exploiting the stationarity principle \cite{huelga1995proposed, huelga1996temporal,  emary2013leggett, Naikoo:2017fos, JavidLGImeson}, which, apart from other conditions demands that if the neutrino is prepared in state $n$ at time $t=0$, then the conditional probabilities $P(n,t+\tau | n,\tau)$ are invariant under   time-translation, i.e., $P(n,t+\tau | n,\tau) = P(n,t|n,0)$. The inequality so obtained could be called entropic Leggett-Garg type inequality, in  consonance with its Leggett-Garg counterparts \cite{Naikoo:2017fos}. From now on, to avoid complexity of notation, we will address the entropic Leggett-Garg type inequality as ELGI.\par
\begin{figure*}[ht]
	\centering
	\includegraphics[width=80mm]{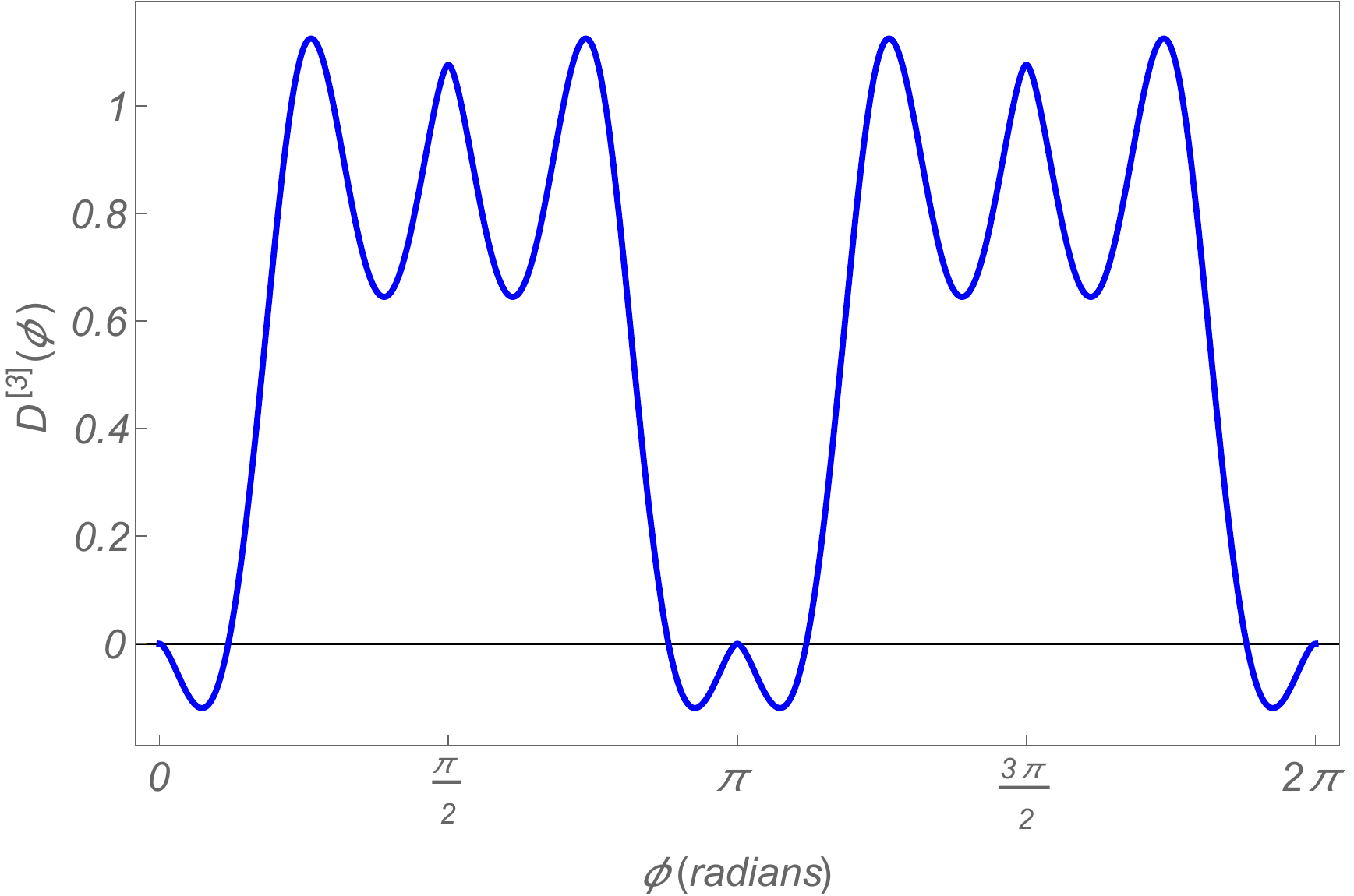}
	\caption{(Color online) Information deficit $\mathcal{D}^{[3]}(\phi)$ plotted against dimensionless parameter $\phi(=(\frac{\Delta_{21} L}{2\hbar c E})$ for \textit{two flavor} approximation in vacuum and three measurements made at $t_0, t_1$ and $t_2$ ($t_0 <t_1 < t_2$). The negative values of  $D^{[3]}(\phi)$ correspond to the violation of ELGI. The values of the mixing angle $\theta_{12}$ and mass squared difference $\Delta_{21}$ are chosen to be  $33.48^{o}$ and $7.5 \times 10^{-5} eV^{2}$, respectively. The result is independent of the initial state chosen, since the survival and oscillation probabilities have  same form irrespective of the initial state. The maximum negative value (measure of the strength of violation) acquired by  $\mathcal{D}^{[3]}(\phi)$ is $-0.1193$.} \label{2F-ELGI}
\end{figure*}
\begin{figure*}[ht!] 
	\centering
	\begin{tabular}{ccc}
		\includegraphics[width=65mm]{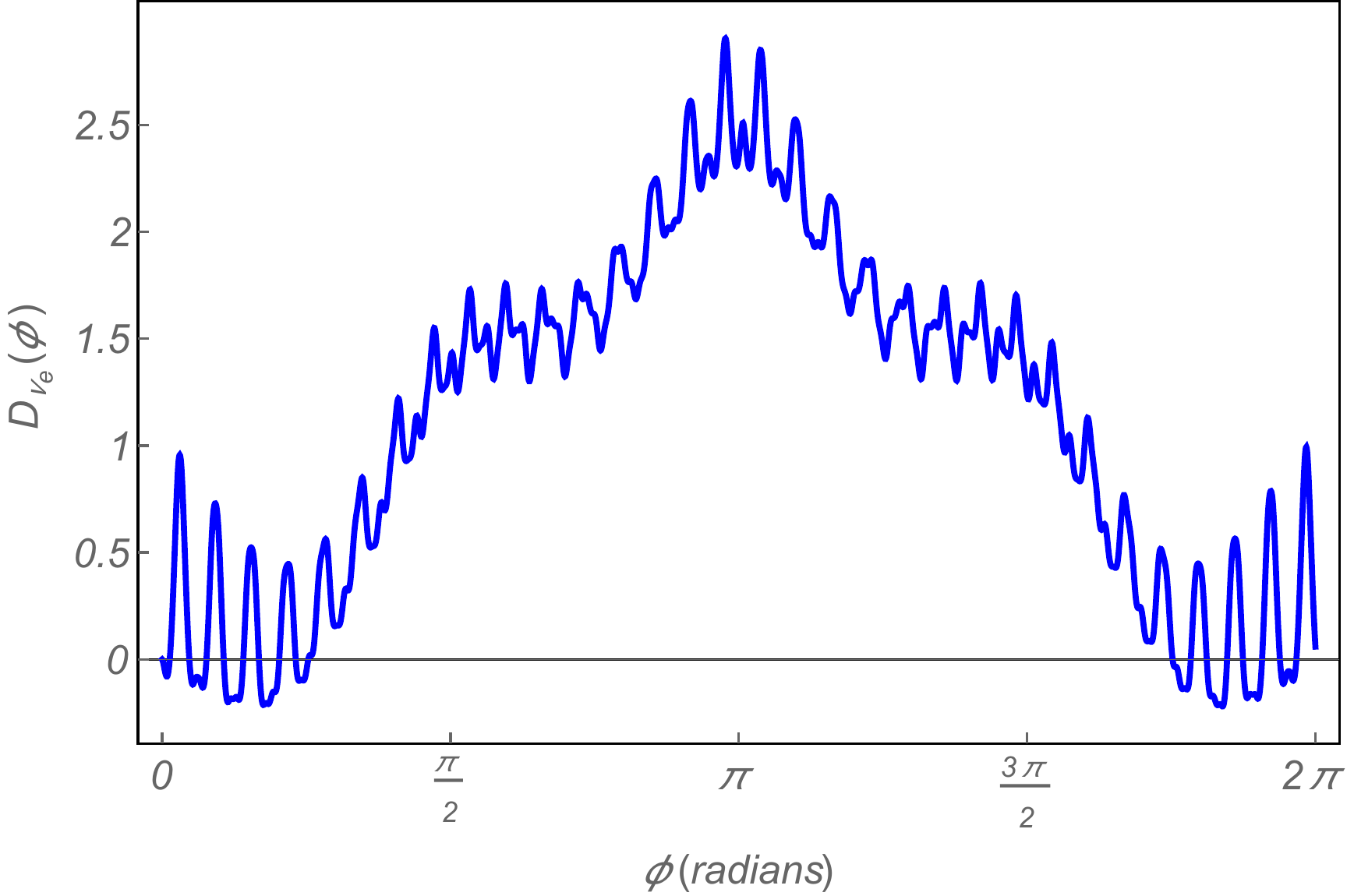}
		\includegraphics[width=65mm]{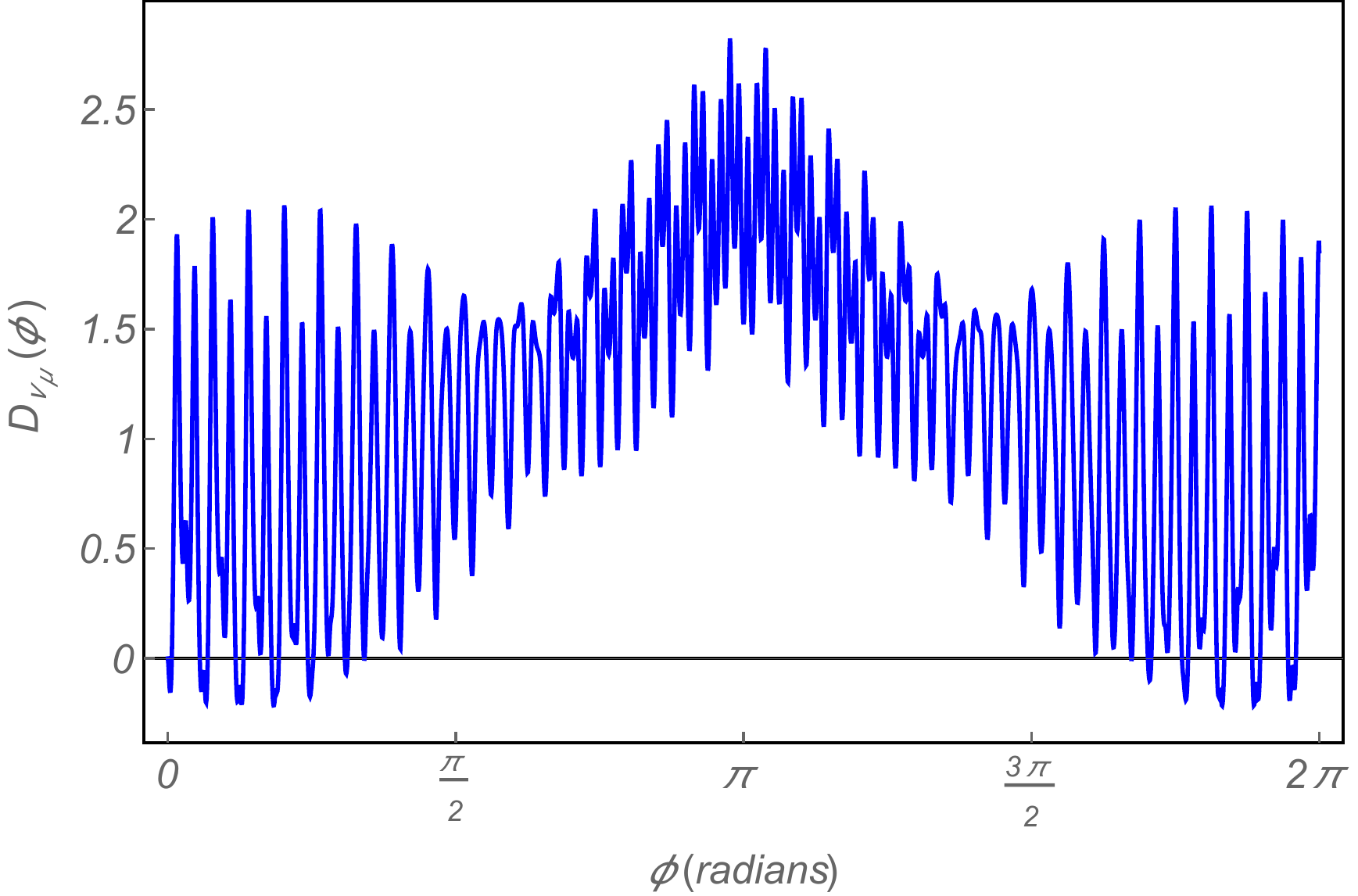}\\
		\includegraphics[width=65mm]{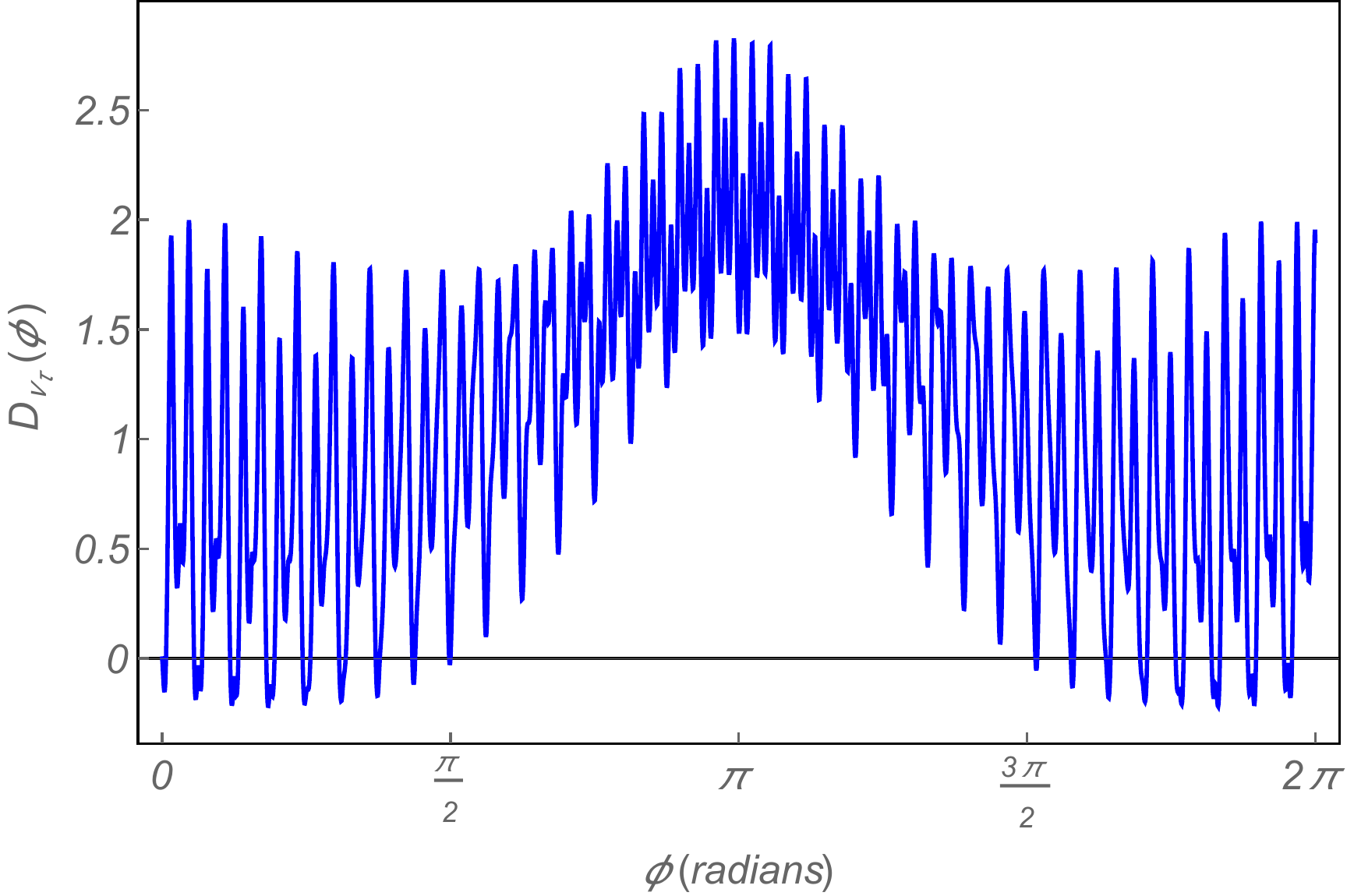} 
	\end{tabular}
	\caption{(Color online) Three flavor scenario in vacuum. Information deficit $\mathcal{D}_{\nu_x}^{[3]}(\phi)( x= e, \mu, \tau)$ plotted against dimensionless parameter $\phi(=(\frac{\Delta_{21} L}{2\hbar c E})$. The various neutrino parameters used are as: $\theta_{12} = 33.48^o,~\theta_{13} = 8.50^o,~\theta_{23} = 42.3^o,~\Delta_{21} = 7.5\times10^{-5}~eV^2,~\Delta_{32} =~\Delta_{31} = 2.457 \times 10^{-3}~eV^2$. The top-left, top-right and bottom figures correspond to the cases with initial state $\nu_e$, $\nu_\mu$ and $\nu_\tau$, respectively. The maximum negative value of the information difference is a measure of the strength of the entropic violation and in this case, turn out to be $Min[\mathcal{D}_{e}^{[3]}(\phi)] \approx -0.2196$ at $\phi \approx 5.7527 ~radians$,  $Min[\mathcal{D}_{\mu}^{[3]}(\phi)] \approx -0.2151$ at $\phi \approx 5.7527 ~radians$,  $Min[\mathcal{D}_{\uptau}^{[3]}(\phi)] \approx -0.2189$ at $\phi \approx 5.7527 ~radians$.}
	\label{3F-ELGI}
\end{figure*}
\begin{figure*}[ht] 
	\centering
	\begin{tabular}{ccc}
		\includegraphics[width=80mm]{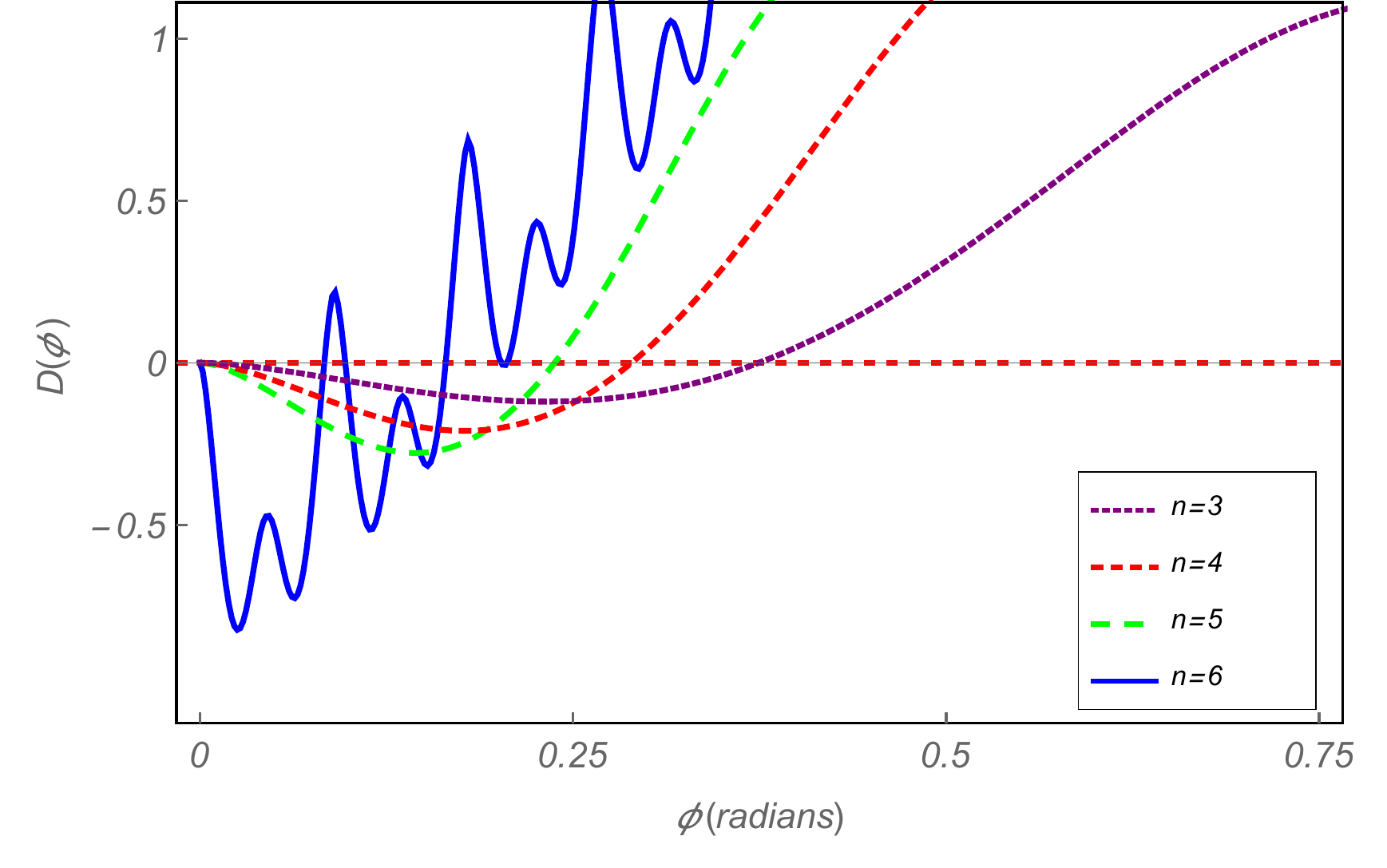} 
		\includegraphics[width=80mm]{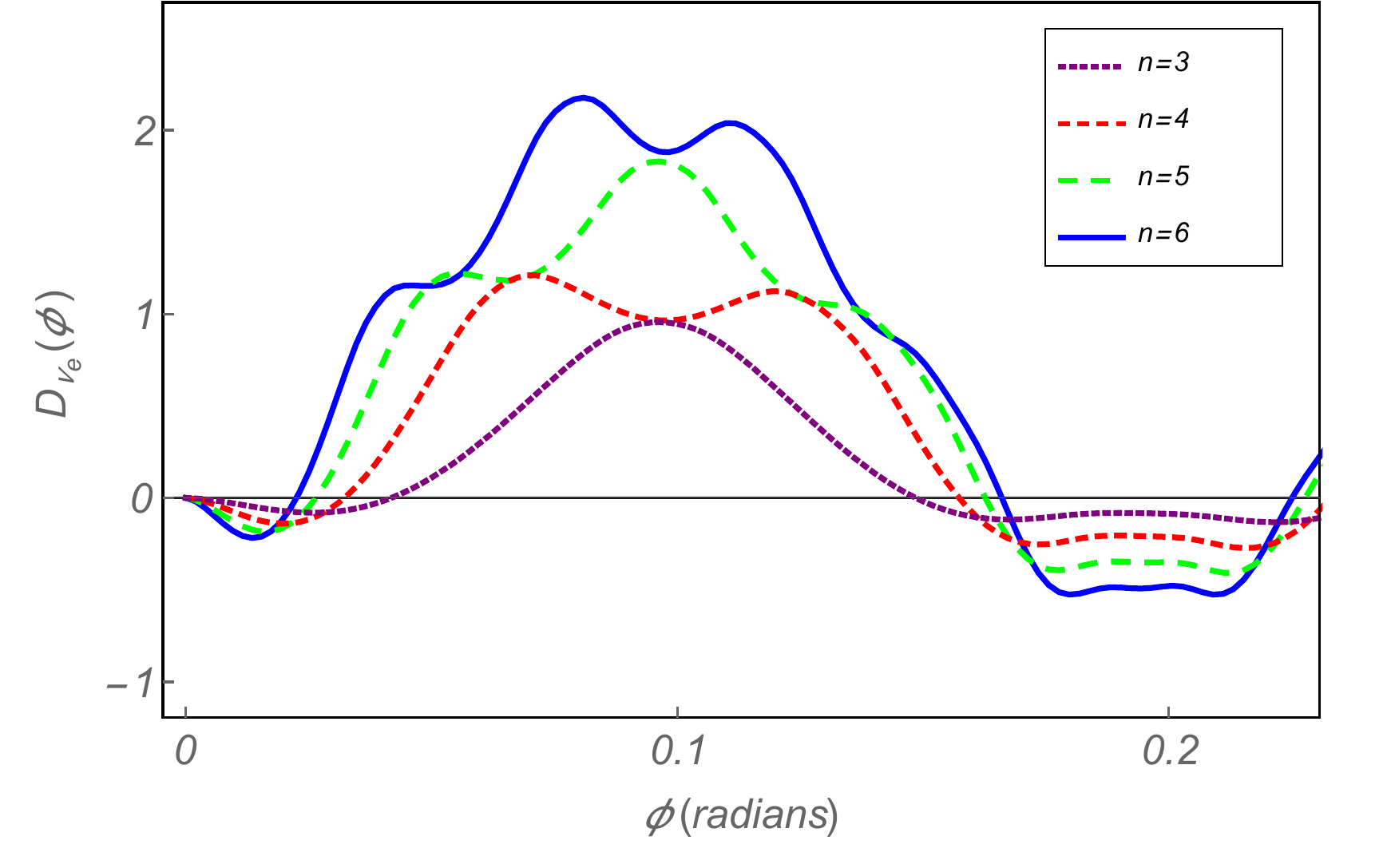} 
	\end{tabular}
	\caption{(Color online) Information difference $\mathcal{D}^{[n]}(\phi)$ plotted against dimensionless parameter $\phi$ for different values of $n$, the number of observations made on the system. The left and right panels correspond to the two and three flavor cases, in vacuum, respectively.  It is clear that, as the number of measurements $n$ increases, the information difference becomes more and more negative. In other words, the maximum negative value of $\mathcal{D}^{[n]}(\phi)$ increases with the increase in the number of measurements. The subscript $\nu_e$ shows that the initial state for the three flavor case is chosen to be $\ket{\nu_e}$.}
	\label{2F-3F-Joint-Zoom}
\end{figure*}
\begin{figure*}[h]
	\centering
		\includegraphics[width=80mm]{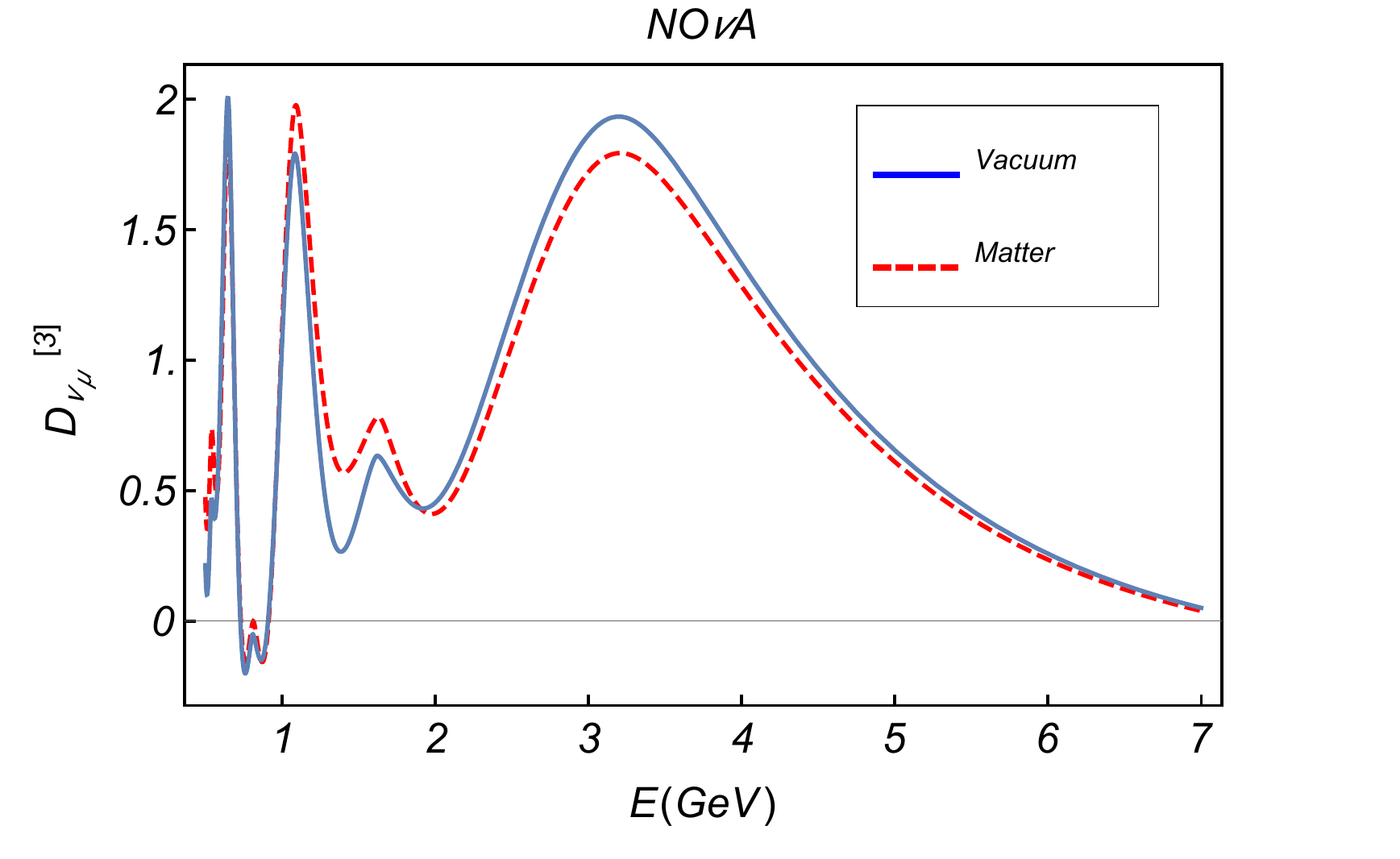}
		\includegraphics[width=80mm]{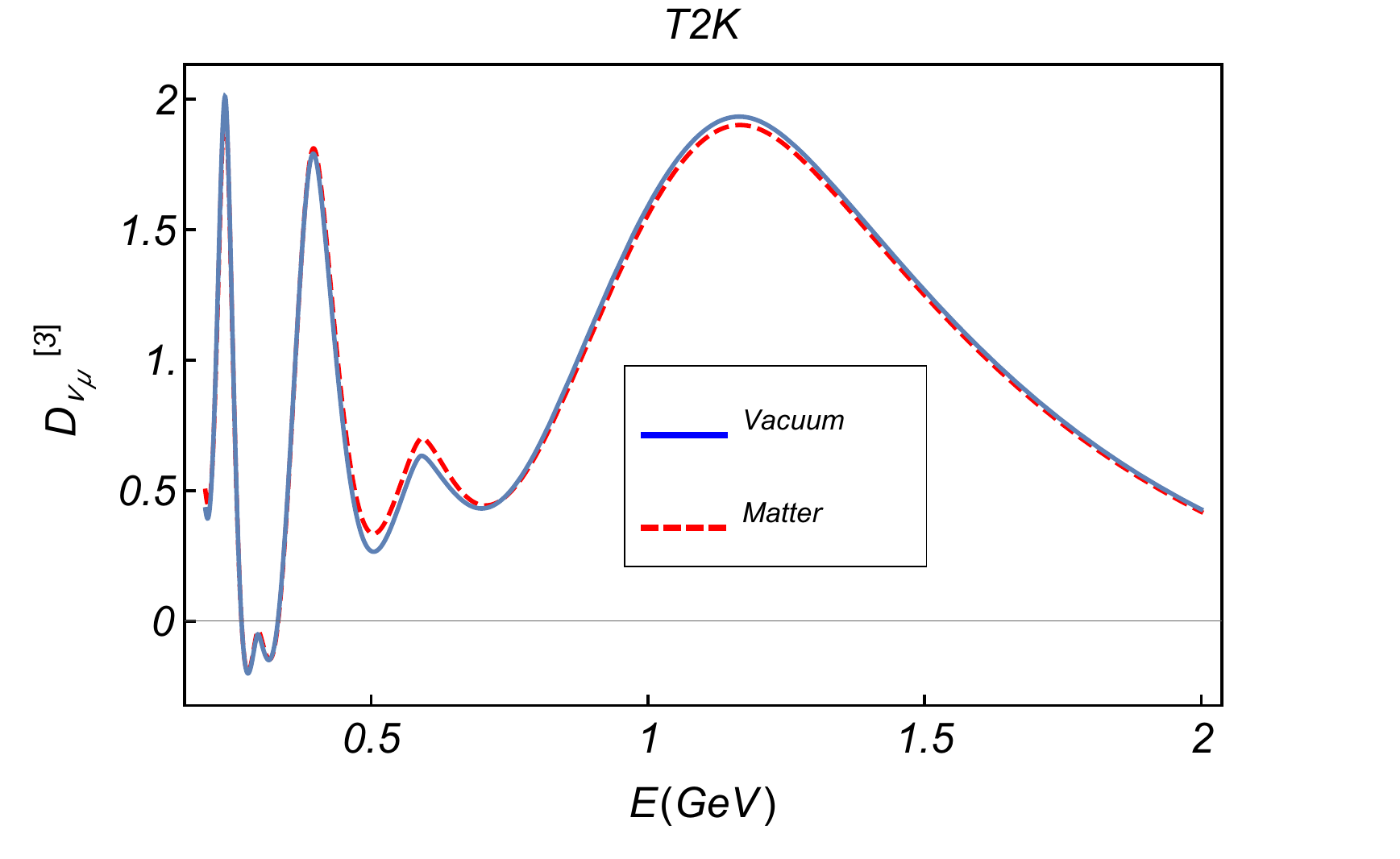}
		\includegraphics[width=80mm]{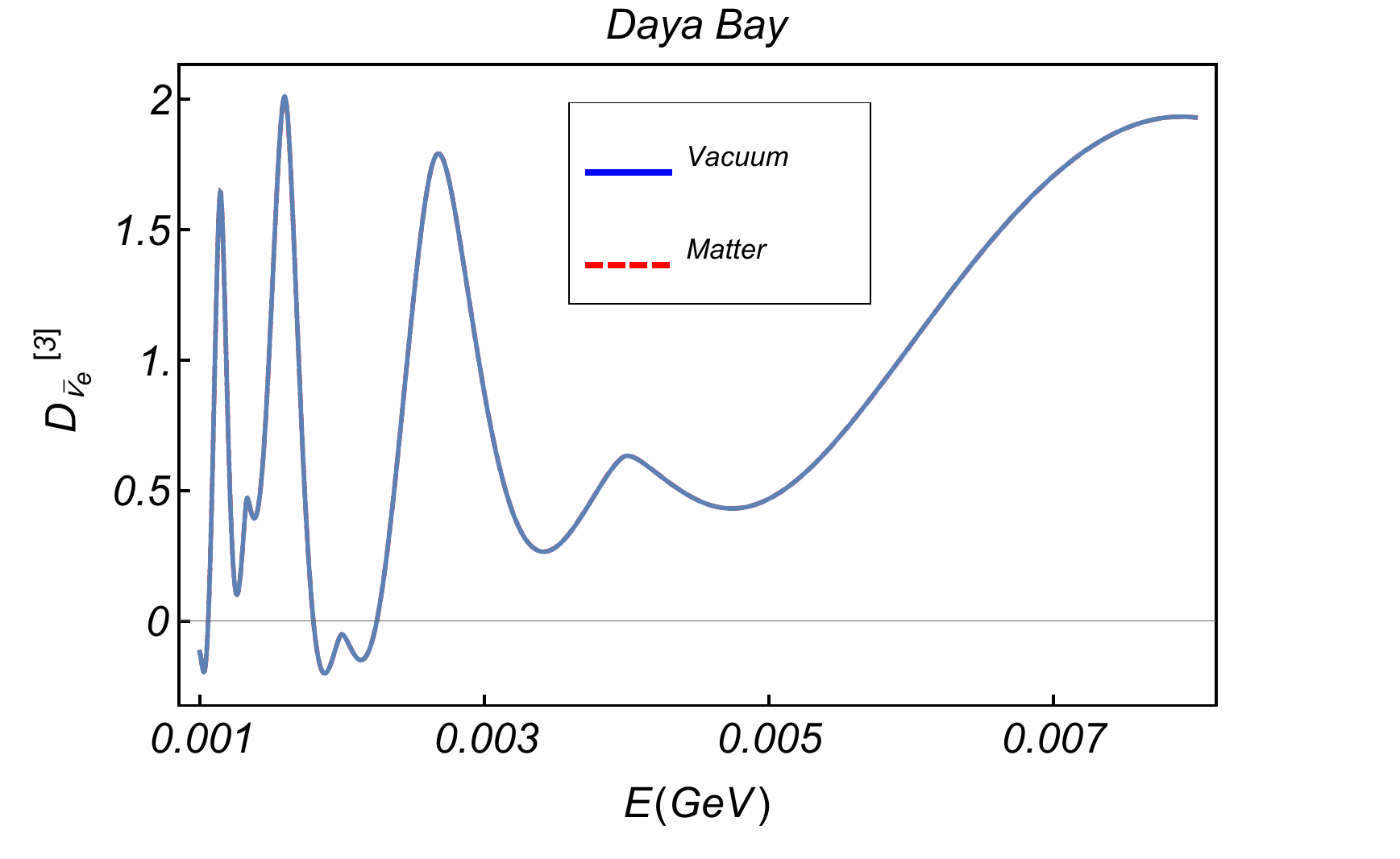}
	\caption{(Color online) Information deficit $\mathcal{D}^{[3]}$ as a function of neutrino energy in three flavor scenario of neutrino oscillation. The top-left, top-right bottom  plots correspond to NO$\nu$A, T2K and Daya-Bay experiments,  respectively. Solid and dashed curves show the variation of $\mathcal{D}^{[3]}$ in matter and vacuum, respectively. The baseline for NO$\nu$A  experiments is 810 km and the energy of the neutrinos varies between 0.5 GeV to 10 GeV. For of T2K experiment, the neutrinos pass through a baseline of 295 km with the energy upto 2 GeV. While as in Daya-Bay experiment, neutrino energy is of the order of few MeVs. It is clear that the matter effect is prominent in long baseline and high energy experiments like NO$\nu$A  than in the small baseline and low energy experiments (T2K and Daya-Bay). The initial flavor in both the accelerator experiments NO$\nu$A and T2K is $\nu_\mu$, while in the reactor Daya-Bay experiment, the initial state is the electron anti-neutrino $\bar{\nu}_e$.} 
	\label{NOvA-T2K}
\end{figure*}

With the notation $H[\nu(t_j)|\nu(t_i)=\nu_{e}](\phi) = H(\phi)$, the ELGI given by Eq. (\ref{Gen-Ineq}), for the neutrino system, under the stationarity assumption discussed above,  becomes
\begin{equation}\label{Deficit}
\mathcal{D}^{[n]}(\phi) = (n-1) H(\phi) - H((n-1)\phi) \ge 0.
\end{equation}
A violation of this inequality, i.e., $\mathcal{D}^{[n]}(\phi) < 0$, would be a signature of the quantum behavior of the system. This information difference is measured in bits ($\log$ to base 2). We have studied this equation for two (Fig. (\ref{2F-ELGI})) and three (Fig. (\ref{3F-ELGI})) flavor scenarios of neutrino oscillations in vacuum. The effect of the number of measurements on the information deficit is depicted in Fig. (\ref{2F-3F-Joint-Zoom}). We also study the  effect of matter density on the deficit parameter in the context of various neutrino experiments as shown in Fig. (\ref{NOvA-T2K}). Discussion of these results  is made in Sec. (\ref{results}).  \par

\FloatBarrier

\section{Time evolution of $B$ and $K$ meson systems}\label{Meson-Dynamics}
In this section we spell out the time evolution of B(K) meson system in the lexicon of open quantum systems. The quantum system is in reality an open system interacting with its  environment. This leads to \textit{decoherence}, the process of loosing the quantum coherence. The study of such decoherence in elementary particles has been a topic of great interest \cite{ELLIS1984381, huet1995violation, ellis1996precision,bertlmann1999quantum,  bertlmann2003decoherence, banerjee2016quantum, Alok:2015iua}.\par
We start by assuming that the Hilbert space is the direct sum $H_{B^o} \bigoplus H_{0}$, spanned by the orthonormal basis  $\ket{B^o} = \begin{bmatrix} 
1 & 0 & 0
\end{bmatrix}^T $, 
$\ket{\bar{B}^o} =  \begin{bmatrix} 
0 & 1 & 0
\end{bmatrix}^T$ and 
$\ket{0}= \begin{bmatrix} 
0 & 0 & 1
\end{bmatrix}^T$, with $\langle B^o | B^o \rangle = \langle \bar{B}^o | \bar{B}^o \rangle =  \langle 0 | 0 \rangle = \mathbb{1}$ and $\langle B^o | \bar{B}^o \rangle = \langle B^o | 0 \rangle = \langle \bar{B}^o | 0 \rangle =  0$. However, the flavor states are not the eigenstates of the time evolution but are related to the stationary states $\{B_H, B_L\}$ by the following equations
\begin{equation}\label{Bo}
\ket{B^o} = \frac{1}{2p} \bigg( \ket{B_L} + \ket{B_H}\bigg); \qquad \ket{\bar{B}^o} = \frac{1}{2q} \bigg( \ket{B_L} - \ket{B_H}\bigg). 
\end{equation}  
Due to normalization $|p|^2 + |q|^2 = 1$ and $\langle B_H | B_L \rangle = |p|^2 - |q|^2$. The existence of CP violation is implied by  $|\frac{p}{q}| \ne 1$.\par                       
The evolution of the system is represented by the operator-sum representation \cite{kraus1983states}
\begin{equation}
\rho(t) = \sum_{i=0}^{5} \mathcal{K}_i(t) \rho(0) \mathcal{K}^{\dagger}_{i}(t),
\end{equation}
where $\mathcal{K}_i(t)$ are the Kraus operators with the following form:
\begin{equation}
\renewcommand{\arraystretch}{1}
\left.\begin{array}{r@{\;}l}
\mathcal{K}_0 =& \ket{0}\bra{0}, \\\\
\mathcal{K}_1 =& \mathcal{C}_{1+}\big( \ket{B^0}\bra{B^0} + \ket{\bar{B}^0}\bra{\bar{B}^0o}\big ) + \mathcal{C}_{1-}\big( \frac{p}{q}\ket{B^0}\bra{\bar{B}^0} + \frac{q}{p}\ket{\bar{B}^0}\bra{B^0} \big),\\\\
\mathcal{K}_2 =& \mathcal{C}_2 \big( \frac{p+q}{2p} \ket{0}\bra{B^0} + \frac{p+q}{2q} \ket{0}\bra{\bar{B}^0} \big), \\\\
\mathcal{K}_3 =& \mathcal{C}_{3+} \frac{p+q}{2p} \ket{0}\bra{B^0} + \mathcal{C}_{3-} \frac{p+q}{2q} \ket{0}\bra{\bar{B}^0} , \\\\
\mathcal{K}_4 =& \mathcal{C}_4 \big( \ket{B^0}\bra{B^0} + \ket{\bar{B}^0}\bra{\bar{B}^0} + \frac{p}{q}\ket{B^0}\bra{\bar{B}^o} + \frac{q}{p}\ket{\bar{B}^0}\bra{B^0} \big),\\\\
\mathcal{K}_5 =& \mathcal{C}_5 \big( \ket{B^0}\bra{B^0} + \ket{\bar{B}^0}\bra{\bar{B}^0} - \frac{p}{q}\ket{B^0}\bra{\bar{B}^0} - \frac{q}{p}\ket{\bar{B}^0}\bra{B^0} \big).
\end{array}\right\} \label{Kraus}
\end{equation}
The coefficients are given by $ \mathcal{C}_{1\pm}=\frac{1}{2} \left[ e^{-(2 i m_L + \Gamma_L + \lambda) t/2} \pm e^{-(2 i m_H + \Gamma_H + \lambda) t/2} \right] $,\\ 	$\mathcal{C}_2  = \sqrt{\frac{Re[\frac{p-q}{p+q}]}{|p|^2 - |q|^2} \big( 1 - e^{-  		\Gamma_L t} - (|p|^2 - |q|^2)^2  \frac{|1 - e^{-(\Gamma + \lambda - i \Delta m )t}|^2}{1 - e^{-\Gamma_H t}}}\big)$, $\mathcal{C}_{3\pm} = \sqrt{\frac{Re[\frac{p-q}{p+q}]}{(|p|^2 - |q|^2)(1 - e^{-\Gamma_H t})}}\big[1 - e^{-\Gamma_H t}  \pm (1 - e^{-(\Gamma + \lambda - i \Delta m)t})(|p|^2 - |q|^2)\big]$, $\mathcal{C}_{4} = \frac{e^{-\Gamma_L t/2}}{2} \sqrt{1 - e^{-\lambda t}}$ and	$\mathcal{C}_{5} = \frac{e^{-\Gamma_H t/2}}{2} \sqrt{1 - e^{-\lambda t}}$. Here $p$ and $q$ are as defined in Eq. (\ref{Bo}). $\Gamma_L$($\Gamma_H$) is the decay width of $B^o_L$($B^o_H$), $\Gamma = 0.5(\Gamma_L + \Gamma_H)$ is the average decay width. $m_H$$(m_L)$ is the mass of $B_H$($B_L$) and $\Delta m = m_H - m_L$ is the mass difference. $\lambda$ is the decoherence parameter which quantifies the strength of the interaction between the one particle system and its environment.\par
If the meson starts at time $t=0$ in state $\ket{B^o}$ or $\ket{\bar{B}^o}$, then at some later time $t$, the state is given by Eqns. (\ref{rhoBt}) and (\ref{rhoBbart}), respectively.
\begin{widetext}
\begin{align}
\rho_{B^0}(t) &= \frac{1}{2}e^{-\Gamma t} \begin{pmatrix}
\cosh(\frac{\Delta  \Gamma}{2}) + e^{-\lambda t} \cos(\Delta m t)                   ~~~& (\frac{q}{p})^* (-\sinh(\frac{\Delta  \Gamma}{2}) - i e^{- \lambda t} \sin(\Delta m t)     &       0 \\\\
(\frac{q}{p}) (-\sinh(\frac{\Delta  \Gamma}{2}) + i e^{-\lambda t} \sin(\Delta m t))  ~~~& |\frac{q}{p}|^2 \cosh(\frac{\Delta  \Gamma}{2}) - e^{-\lambda t} \cos(\Delta m t)         &       0  \\\\
0                                       ~~~&                0                                      &     \rho_{33}(t)
\end{pmatrix} \label{rhoBt},\\ \nonumber \\ 
\rho_{\bar{B}^0}(t) &= \frac{1}{2} e^{-\Gamma t} \begin{pmatrix}
|\frac{p}{q}|^2 (\cosh(\frac{\Delta  \Gamma}{2}) - e^{- \lambda t} \cos(\Delta m t))      ~~&   (\frac{p}{q}) (-\sinh(\frac{\Delta  \Gamma}{2}) + i e^{- \lambda t} \sin(\Delta m t))  &  0  \\\\
(\frac{p}{q})^* (-\sinh(\frac{\Delta  \Gamma}{2}) -i e^{- \lambda t} \sin(\Delta m t))    &    \cosh(\frac{\Delta  \Gamma}{2}) + e^{- \lambda t} \cos(\Delta m t)                      &  0   \\\\
0                                             &           0                                          &  \tilde{\rho}_{33}(t) \\
\end{pmatrix}  \label{rhoBbart},
\end{align}
\end{widetext}
where $\Delta \Gamma = \Gamma_H - \Gamma_L$, $\rho_{33}(t)$ and $\tilde{\rho}_{33}(t)$  are complicated functions. The same description holds for the case of $K$ meson system with appropriate notational changes. The diagonal elements give survival and transition probabilities
\begin{align}
P_{B^o \rightarrow B^o}(t) &= \frac{1}{2} e^{-\Gamma t}\bigg[\cosh(\frac{\Delta  \Gamma}{2}) + e^{-\lambda t} \cos(\Delta m t) \bigg], \\
P_{B^o \rightarrow \bar{B}^o}(t) &= \frac{1}{2} e^{-\Gamma t}\bigg[ |\frac{q}{p}|^2 \cosh(\frac{\Delta  \Gamma}{2}) - e^{-\lambda t} \cos(\Delta m t) \bigg].
\end{align} 
Similarly, we can define probabilities like $P_{\bar{B}^o \rightarrow B^o}(t)$, $P_{\bar{B}^o \rightarrow \bar{B}^o}(t)$, $P_{\bar{B}^o \rightarrow 0}(t)$.\par
\begin{figure*}[ht] 
	\centering
		\includegraphics[width=80mm]{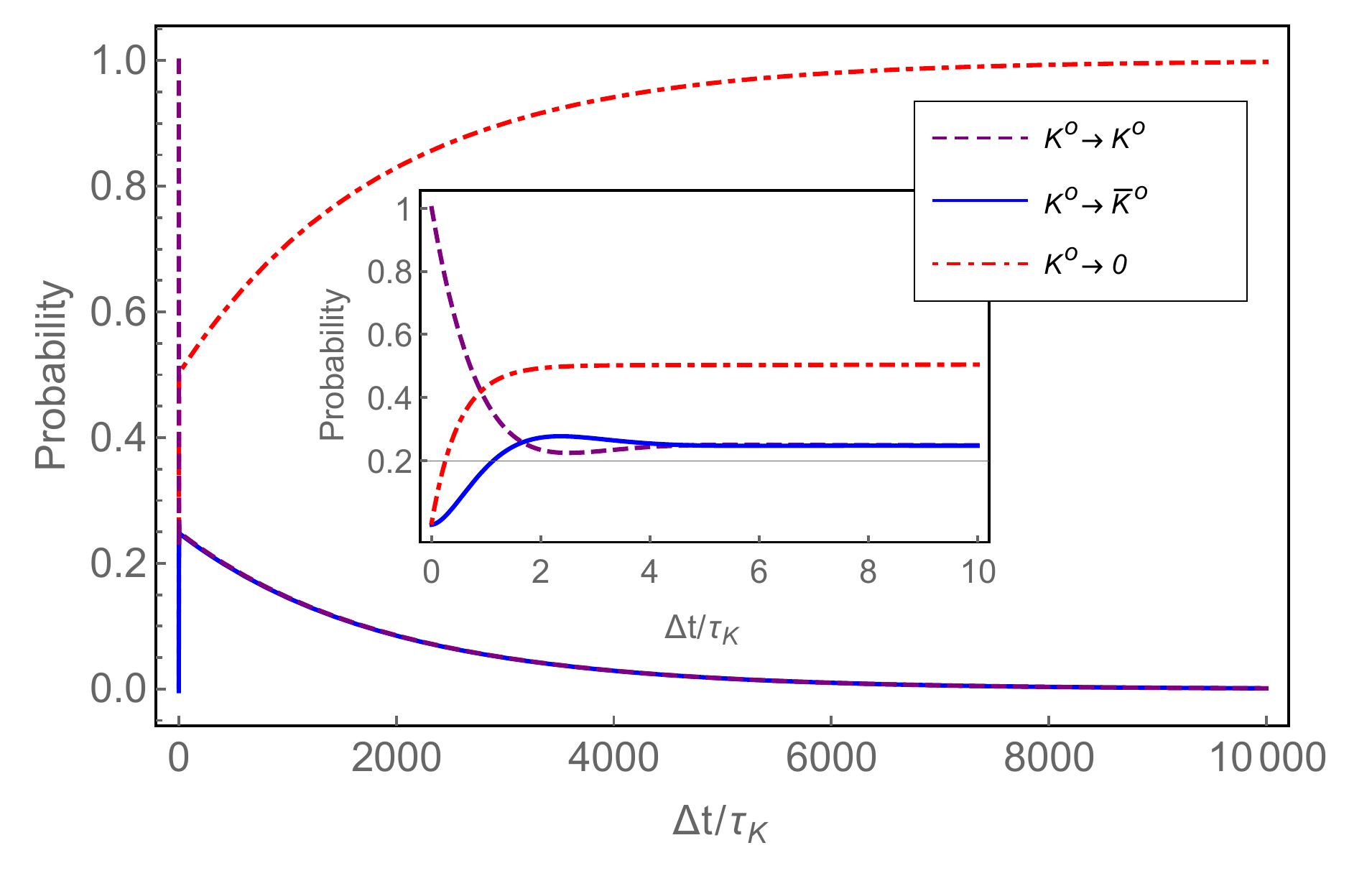}
		\includegraphics[width=80mm]{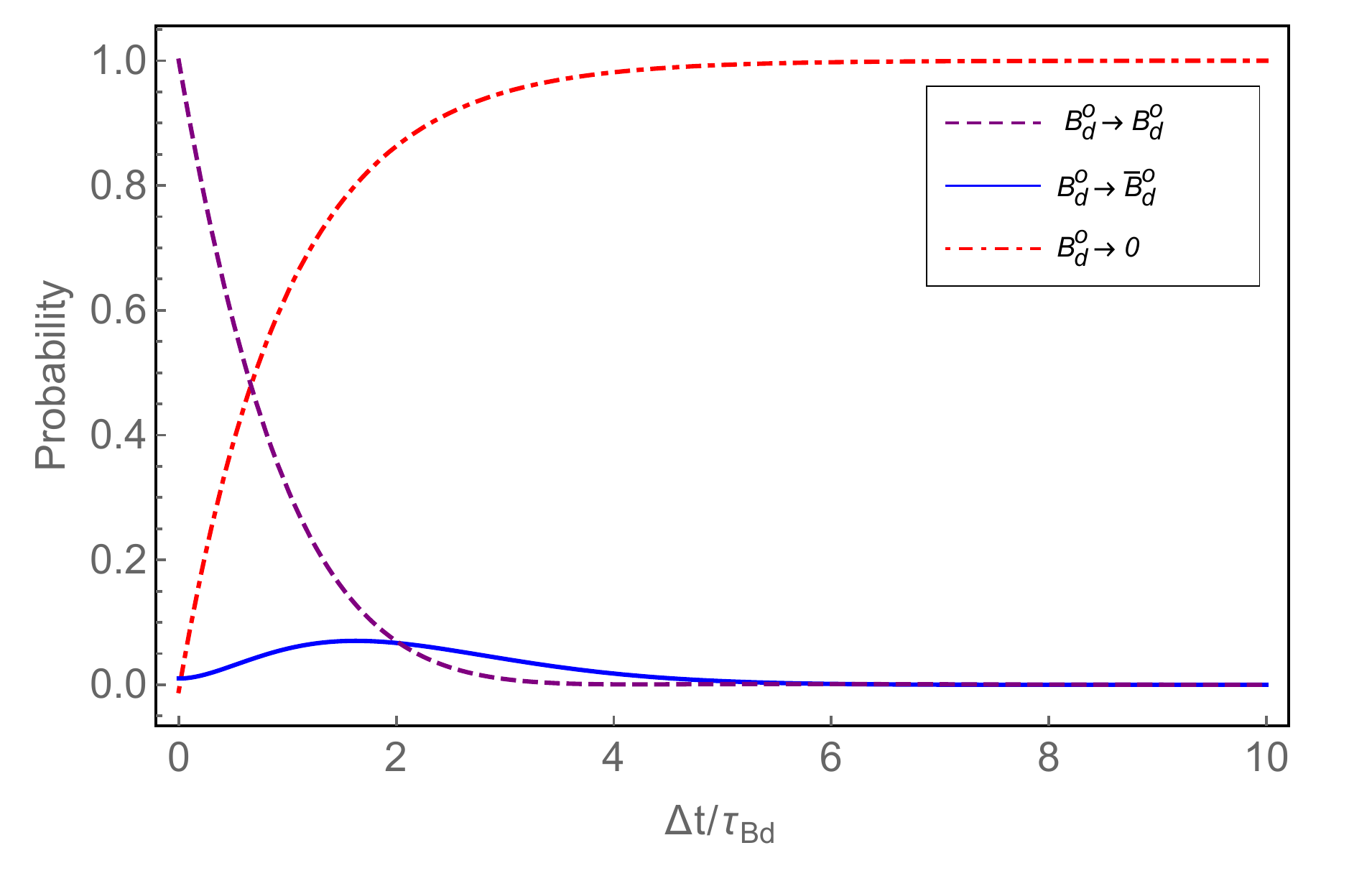}
		\includegraphics[width=80mm]{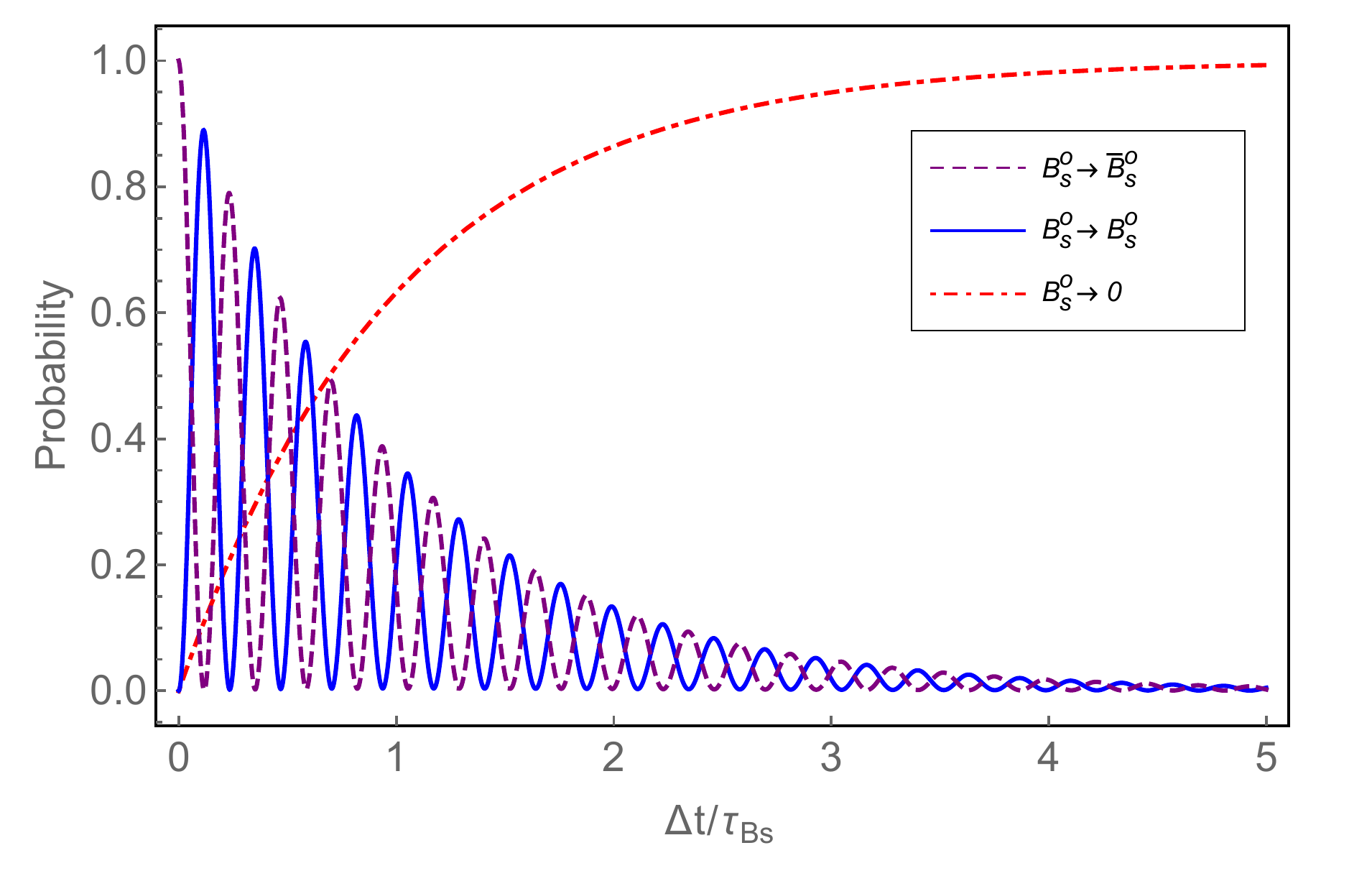} 
	\caption{(Color online) Probabilities: The  top-left, top-right and bottom figures depict the probabilities plotted {\it w.r.t} the dimensionless quantity $\Delta t/\uptau$ for the $K$, $B_d$ and $B_s$ mesons, respectively.  Here $\Delta t$ is the time between successive measurements and $\tau$ is the mean life time of respective mesons.  For the $K$ system, the mean life time is $  \uptau_K =  1.7889 \times 10^{-10} s $.  Also,  $\Gamma = 5.59 \times 10^{9}~ {\rm s^{-1}}$,  $ \Delta \Gamma = 1.1174 \times 10^{10}~{\rm s^{-1}}$,  $\lambda = 2.0 \times 10^{8}~ {\rm s^{-1}}$ and $\Delta m = 5.302\times 10^{9} ~ {\rm s^{-1}}$ \cite{Olive:2016xmw}. Here we used $ Re(\epsilon) = 1.596 \times 10^{-3} $ and $ |\epsilon| = 2.228 \times 10^{-3} $ \cite{d2006determination}. For the $B_d$ system, $\uptau_{B_d} = 1.518 \times 10^{-12} s $, $ \Gamma = 6.58 \times 10^{11}~ {\rm s^{-1}}$, $\Delta \Gamma = 0$,  $\lambda = 0.012 \times 10^{12} ~{\rm s^{-1}}$ \cite{Alok:2015iua}  and $\Delta m = 0.5064\times 10^{12} ~{\rm s^{-1}}$ \cite{Amhis:2016xyh}. The $CP$ violating parameter used here is $|\frac{q}{p}| = 1.010$ \cite{Amhis:2016xyh}.  Finally, for the $B_s$ meson, $\uptau_{B_s} = 1.509 \times 10^{-12} s $, $ \Gamma = 0.6645 \times 10^{12}~ {\rm s^{-1}}$, $\Delta \Gamma = 0.086 \times 10^{12}~ {\rm s^{-1}}$, $\lambda = 0.012 \times 10^{12}~ {\rm s^{-1}}$ and $\Delta m = 17.757\times 10^{12}~ {\rm s^{-1}}$ \cite{Amhis:2016xyh}. The value of the $CP$ violating parameter here is $|\frac{q}{p}| = 1.003$ \cite{Amhis:2016xyh}.}\label{prob}
\end{figure*}

Based on the above discussion, we show in Fig. (\ref{prob})  the probabilities for the case of K and B meson systems. The sub-figure, in case of $K$ system, highlights a region between $\Delta t = 0~to~10 \tau_K$. It can be seen that the K meson system retains coherence for much longer time, consistent with the findings reported in \cite{banerjee2016quantum}.

\FloatBarrier

\textit{Entropic Leggett-Garg inequality for $B$ and $K$ meson systems:}\\
In this case, the joint probability  $P(a_{2},a_1)$ is constructed in terms of the Kraus operators defined in Eq. (\ref{Kraus})  as 
\begin{equation}
P(a_{2},a_1) = Tr\Big\{\Pi^{a_2} \sum\limits_{i=0}^{5} \mathcal{K}_{i} (t_2 - t_1)\Pi^{a_1} \rho(t_1) \mathcal{K}^{\dagger}_{i}(t_2 - t_1) \Big\},
\end{equation}
with $\Pi^{a_i} = |M \rangle \langle M |_{t_i}$ is the projector corresponding to the measurement at time $t_i$. In particular, if the meson is produced in flavor state $\ket{B^o}$ at time $t_0$ and a measurement is made at a later time $t_1$, then the mean conditional entropy, in terms of various probabilities is given by

\begin{align}
H[A(t_1)|A(t_0) = B^o] &=  -P_{B^o \rightarrow B^o}(\Delta t) \log_2 P_{B^o \rightarrow B^o}(\Delta t) \nonumber \\&- 
P_{B^o \rightarrow \bar{B}^o}(\Delta t) \log_2 P_{B^o \rightarrow \bar{B}^o}(\Delta t)\nonumber \\&- 
P_{B^o \rightarrow 0}(\Delta t) \log_2  P_{B^o \rightarrow 0}(\Delta t).  
\end{align}

Here we have used $\Delta t = t_1 - t_0$. Similarly we can calculate $H[A(t_2)|A(t_1)]$ and $H[A(t_2)|A(t_0)]$ and construct the simplest ELGI  by invoking the stationarity assumption discussed in Sec. (\ref{Neutrino-Osc}), such that the $n$-measurement inequality reads
\begin{equation}\label{Deficit-Meson}
\mathcal{D}^{[n]}(\Delta t) = (n-1) H(\Delta t) - H((n-1)\Delta t)\ge0.
\end{equation} 
Hence, a violation of ELGI would imply negative values of $\mathcal{D}^{[n]}(\Delta t)$.  We now discuss the various results obtained by studying the ELGI for neutrino and meson systems.

\section{Results and discussion}\label{results}
Figure (\ref{2F-ELGI}) shows the variation of the information deficit  $\mathcal{D}^{[3]}(\phi)$ in two flavor approximation for $n=3$, the number of measurements made on the system. Since the survival and oscillation probabilities, in two flavor approximation of neutrino oscillation, are independent of the initial state, so is the information difference $\mathcal{D}^{[3]}(\phi)$. The maximum negative value of $\mathcal{D}^{[3]}(\phi)$ is $-0.1193$.  Fig. (\ref{3F-ELGI}) depicts the same for three flavor case with different initial states. A clear violation of the ELGI is seen in all the cases. We find that the maximum negative value of   $\mathcal{D}^{[3]}(\phi)$(measure of  the strength of entropic violation) is same upto second decimal place, Min$[\mathcal{D}^{[3]}(\phi)] \approx -0.21$ occuring at $\phi \approx 5.75~radians$. Thus, the strength of violation  in three flavor case in approximately twice as that of the two flavor scenaio, i.e., Min$[\mathcal{D}^{[3]}(\phi)]~(3-flavor) \approx 1.75$ and Min$[\mathcal{D}^{[3]}(\phi)~(2-flavor)]$. The maximum negative value for $\mathcal{D}^{[3]}(\phi)$ increases as we increase $n$-the number of measerements, as shown in Fig. (\ref{2F-3F-Joint-Zoom}). A similar trend was observed in \cite{devi2013macrorealism} for a quantum spin-$s$ system.\par
So far we discussed an ideal scenario of neutrinos propagating in vacuum and also talked about the $n$ time measurements. From the experimental point of view, the neutrinos do interact with matter, although the interaction is quite feeble. Also the possibility of putting up multiple detectors and making arbitrary number of measurements is difficult, given the present experimental facilities. Therefore, it is interesting to study the ELGtI in the context of some ongoing experiments. We will now discuss the violation of ELGtI by taking inputs parameters (viz., energy of neutrino, baseline and matter density) from the experiments like NO$\nu$A, T2K and Daya-Bay.\par
For the case when neutrinos pass through a constant matter density, one can obtain analytic form of the time evolution operator both in the flavor and mass basis \cite{Ohlsson:1999um,ohlsson2000three,Ohlsson:2001et}. In flavor basis, the time evolution operator $\mathcal{U}_f$ takes a state $\psi(t_1) $ at  time $t_1$ to $\psi(t_2)$ at some late time $t_2$  such that $\psi(t_2) = \mathcal{U}_f(t_2 - t_1) \psi(t_1)$. In the ultra-relativistic limit $t \approx L$, where $L$ is the distance traveled by the neutrino. Apart from $L$, the time evolution operator depends on the energy of the neutrino $E_n$, the $CP$ violating phase $\delta$, the matter density parameter $A=\pm \sqrt{2} G_F N_e$  ($G_F$ is the Fermi coupling constant, $N_e$ is the electron density of the medium), the mixing angles ($\theta_{13}$, $\theta_{23}$, $\theta_{13}$) and the mass square differences ($\Delta_{21}$, $\Delta_{31}$, $\Delta_{32}$). Fig. (\ref{NOvA-T2K}) shows the variation of the deficit parameter $\mathcal{D}^{[3]}$ with the energy of the neutrinos for accelerator experiments like NO$\nu$A and T2K and the reactor  Daya-Bay experiment in vacuum (red dashed) and matter (solid blue). It can be seen that the matter effect is more prominent in the long baseline and high energy experiment NO$\nu$A than the relatively short baseline and low energy experiments like  T2K and Daya-Bay.

\begin{figure*}[ht] 
	\centering
		\includegraphics[width=80mm]{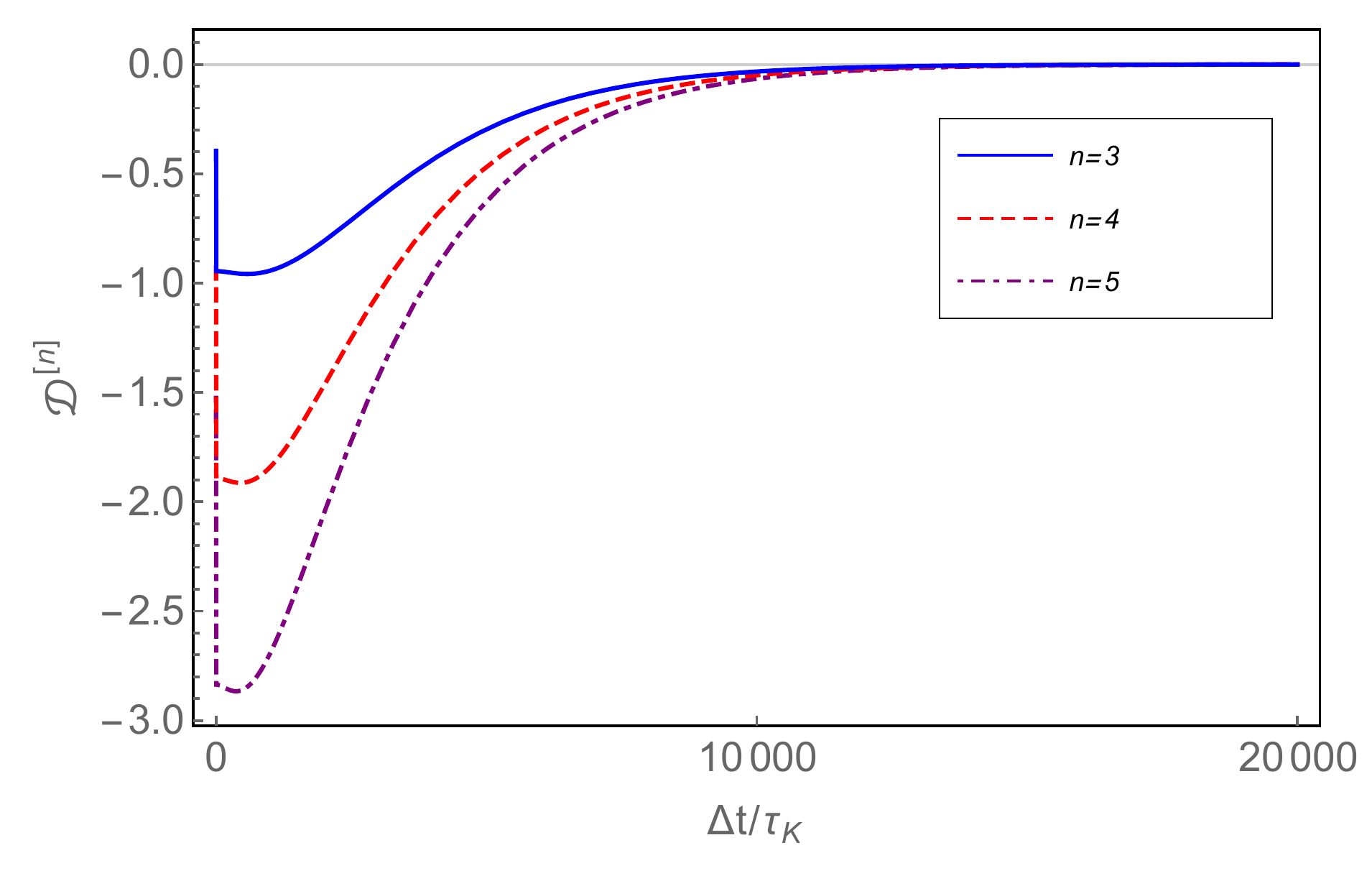}
		\includegraphics[width=80mm]{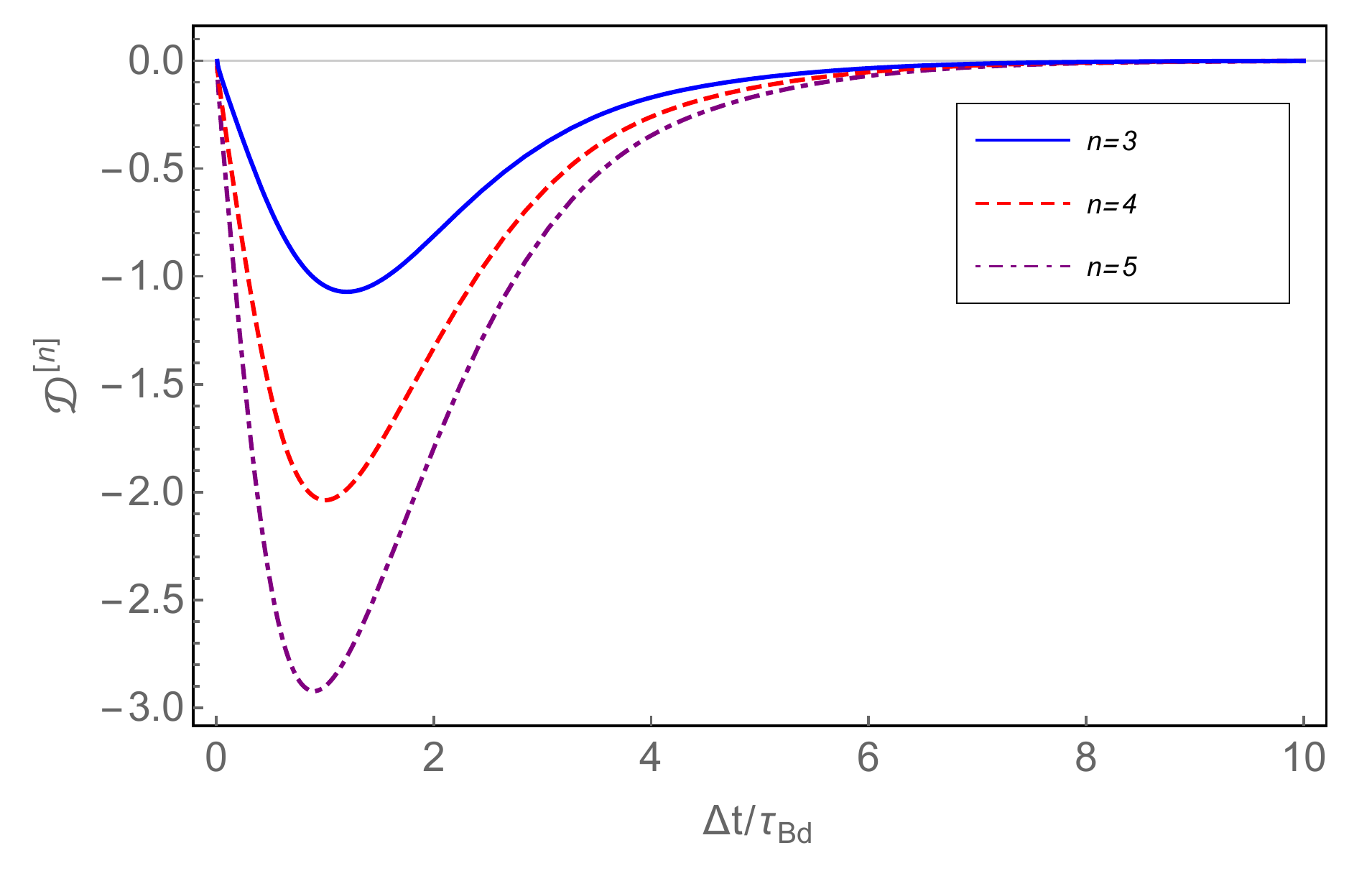}
		\includegraphics[width=80mm]{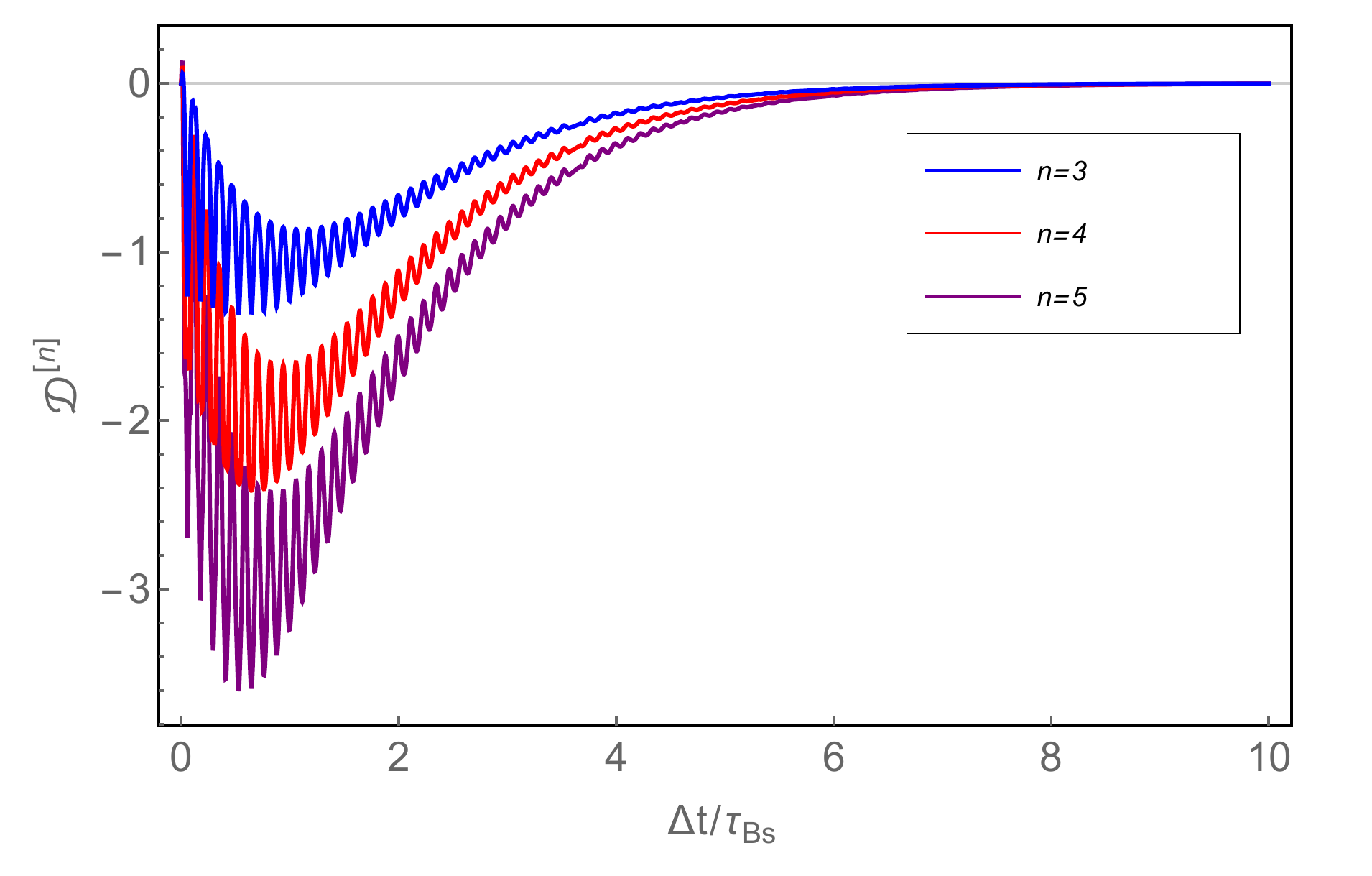} 
	\caption{(Color online) Information deficit parameter $\mathcal{D}^{[n]}$ plotted against the dimensionless quantity $\Delta t/ \uptau $, for different number of measurements $n$. The various parameters used are the same as used in Fig. (\ref{prob}).}\label{fig-deficit}
\end{figure*}

Figure (\ref{prob}) depicts the survival and transition probabilities for the decohering neutral  $K$, $B_d$ and $B_s$ meson systems. The mean-lifetime of the neutral $K$ meson system is much longer than its $B$ meson counterpart.  The information deficit is plotted in Fig. (\ref{fig-deficit}) for $K$, $B_d$ and $B_s$ systems. It is clear that the ELGI for three time measurement is violated in all the three cases. The extent of violation increases with the increase in the number of measurements $n$. Also, the time for which $\mathcal{D}^{[3]}$ remains negative (before it touches the classical limit 0), also increases with the increase in  number of measurements. Analogous features were seen in \cite{devi2013macrorealism} in the context of a spin-$s$ system. The violation sustains for a much longer time in $K$ meson system than in $B_d$ and $B_s$ systems, bringing out the point that the $K$ meson system sustains its quantum behavior for a much longer time as compared to $B$ meson system, consistent with  earlier works. The oscillatory behavior of $\mathcal{D}^{[3]}$, in the $B_s$ system,  is because of the fact that the mass difference $\Delta m$ for $B_s$ system is nearly 35 times the value for the  $B_d$ system and plays the role of frequency, in the form of terms like $\cos\Delta mt (\sin\Delta m t)$, in the state matrix and hence in the probabilities.\par

\begin{figure*}[h] 
	\centering
		\includegraphics[width=80mm]{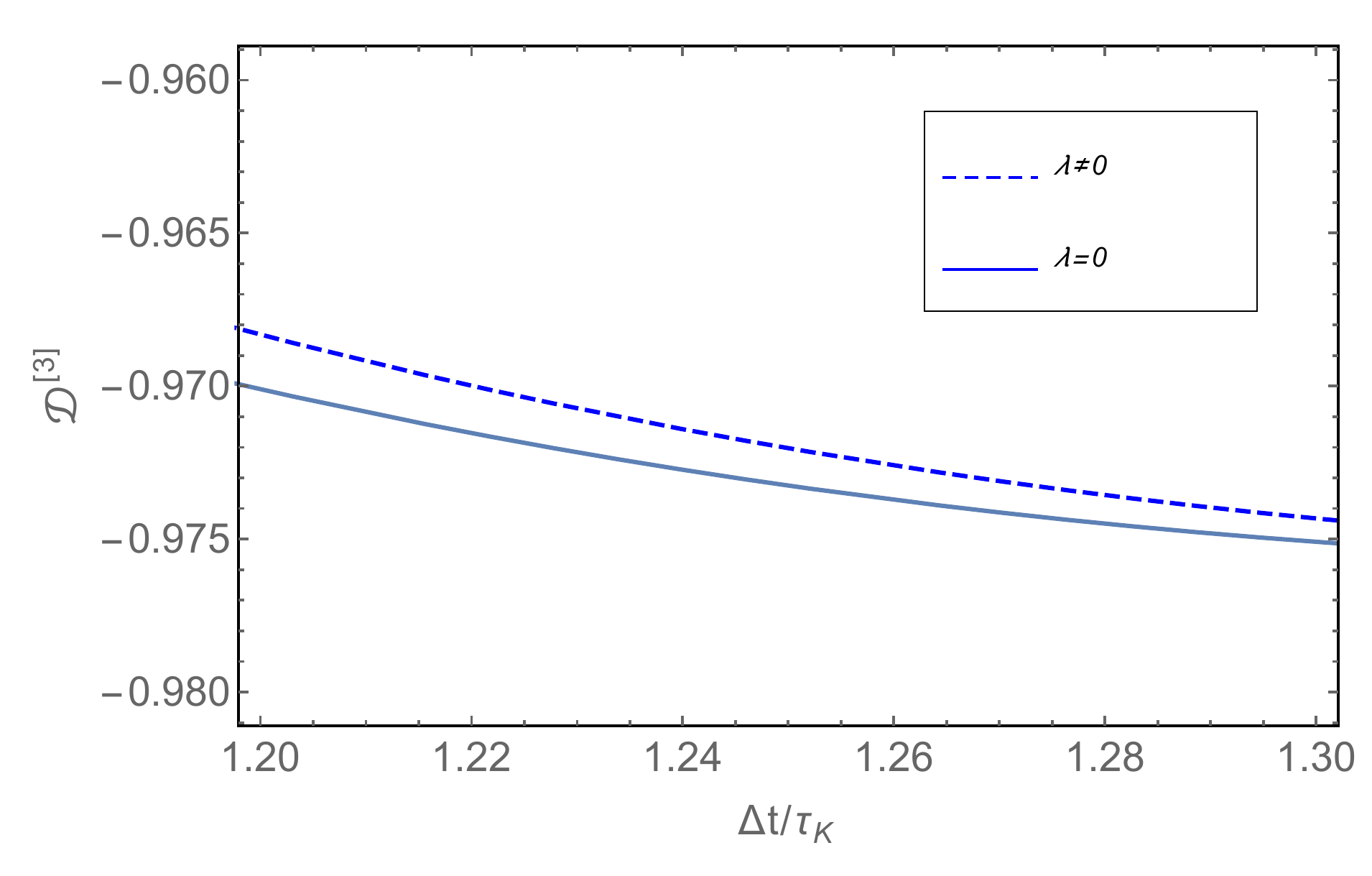}
		\includegraphics[width=80mm]{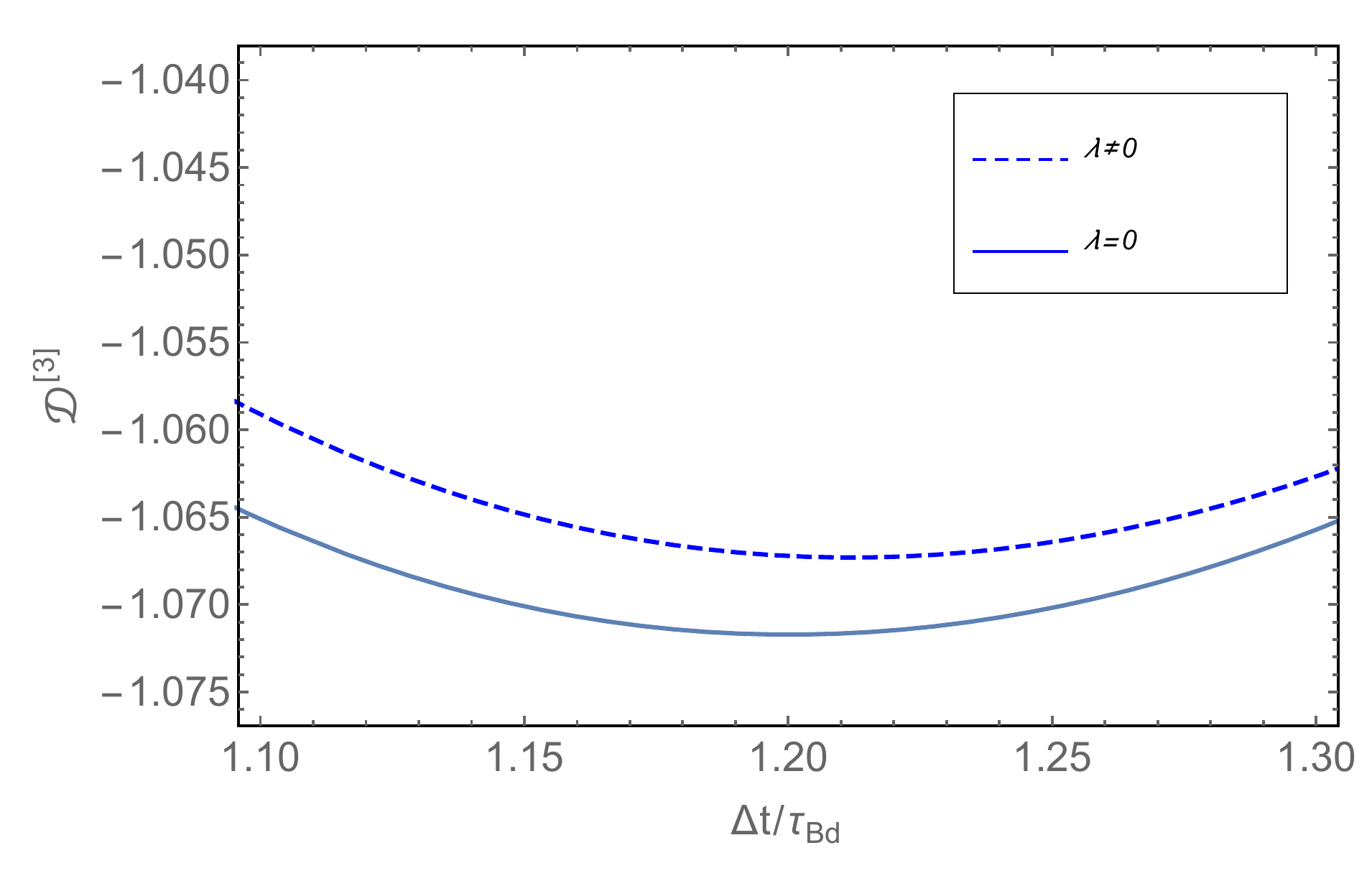}
		\includegraphics[width=80mm]{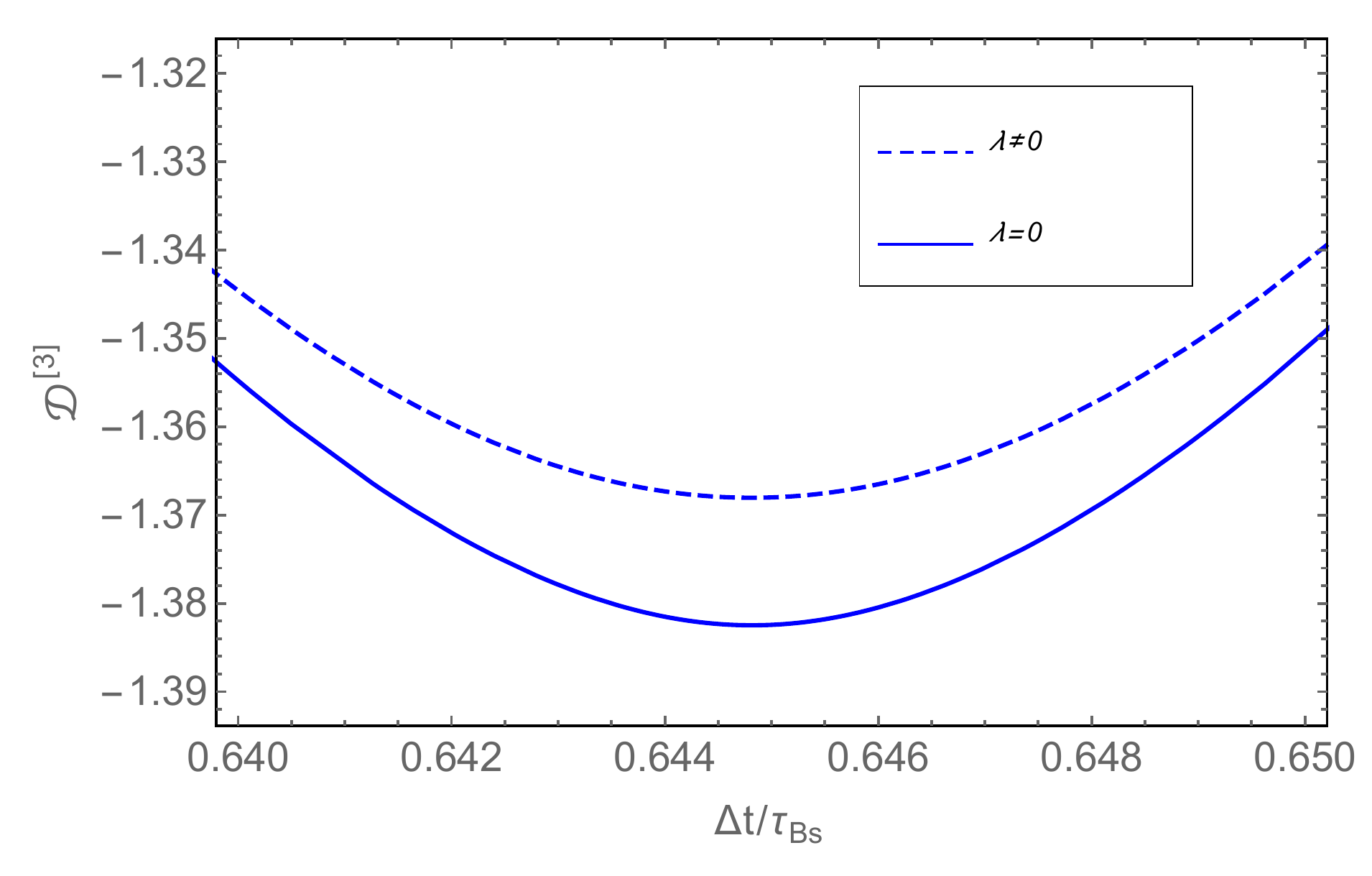} 
	\caption{(Color online) Information deficit parameter $\mathcal{D}^{[3]}$  with and without decoherence. It can be seen that the non-zero value of the decoherence parameter $\lambda$ brings $\mathcal{D}^{[3]}$ closer to its classical limit, zero.}\label{fig-decoherence}
\end{figure*}
Decoherence is the process of loosing quantum coherence. In other words, the system comes close to the classical domain. It can be seen from Fig. (\ref{fig-decoherence}) that the effect of decoherence is to bring the deficit parameter $\mathcal{D}^{[3]}$ closer to the classical value zero, as expected.\par 
  The ELGI for neutrino and meson systems given by Eqs. (\ref{Deficit}) and (\ref{Deficit-Meson}), respectively, are in terms of measurable quantities, i.e., the \textit{survival} and \textit{transition} probabilities. Several neutrino experiments like NO$\nu$A, T2K, Daya-Bay, use a neutrino source producing neutrinos in a particular state  $\nu_\mu$ and the detector is sensitive to detect a particular flavor state $\nu_e$.  Therefore, by using the experimentally observed probabilities at various energies, one can compute the information deficit parameter, thereby verifying the ELGI. For meson system, the state of a neutral meson is determined using the method of \textit{tagging}. This allows one to determine the survival and transition probability of the neutral meson by identifying the charge of the lepton in its semileptonic decay. A knowledge of these probabilities would allow one to compute the deficit parameter and hence verify the ELGI.

\section{Conclusion}\label{conclusion}
\begin{figure*}[h!]
	\centering
	\includegraphics[width=80mm]{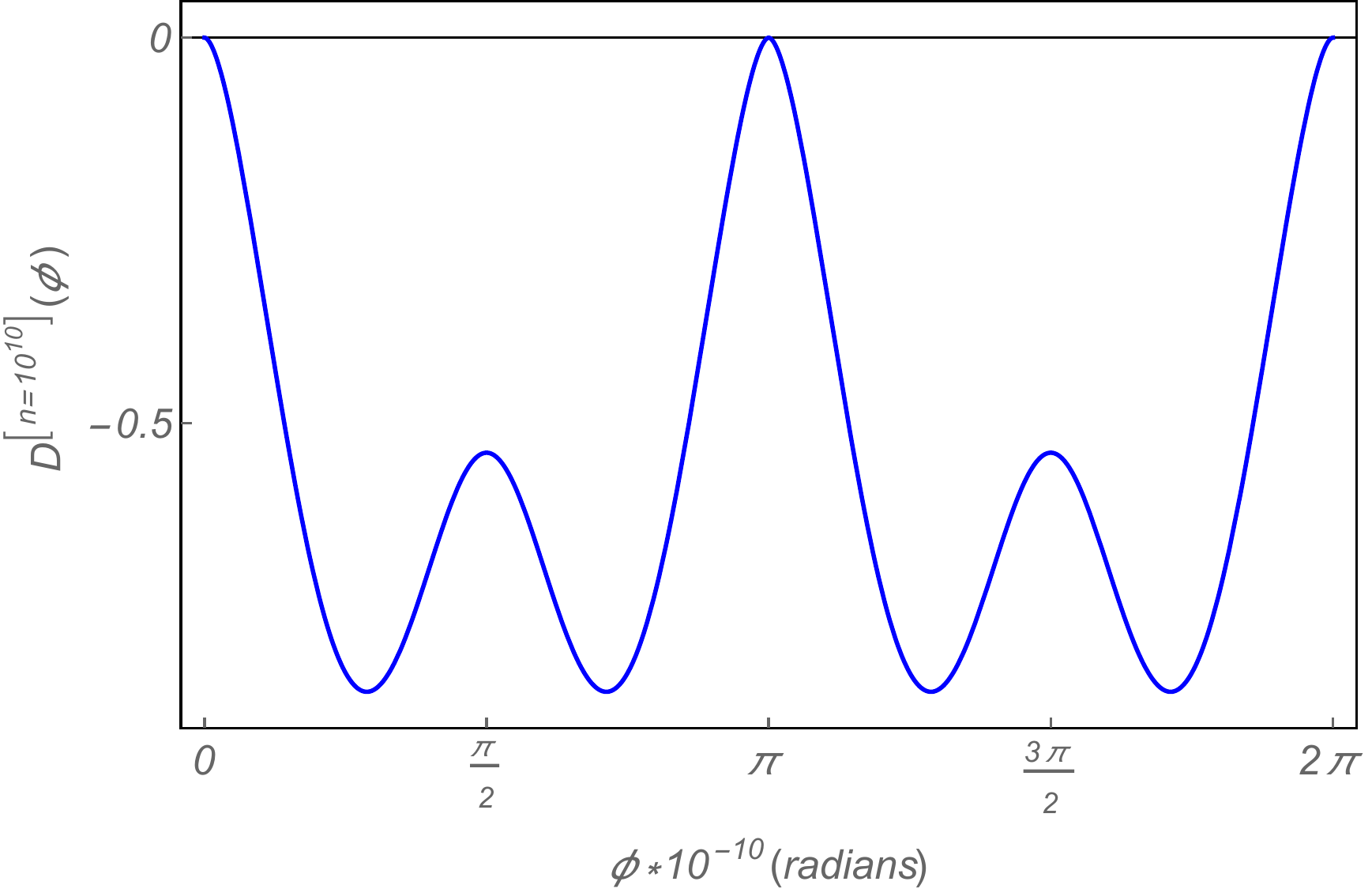}
	\includegraphics[width=80mm]{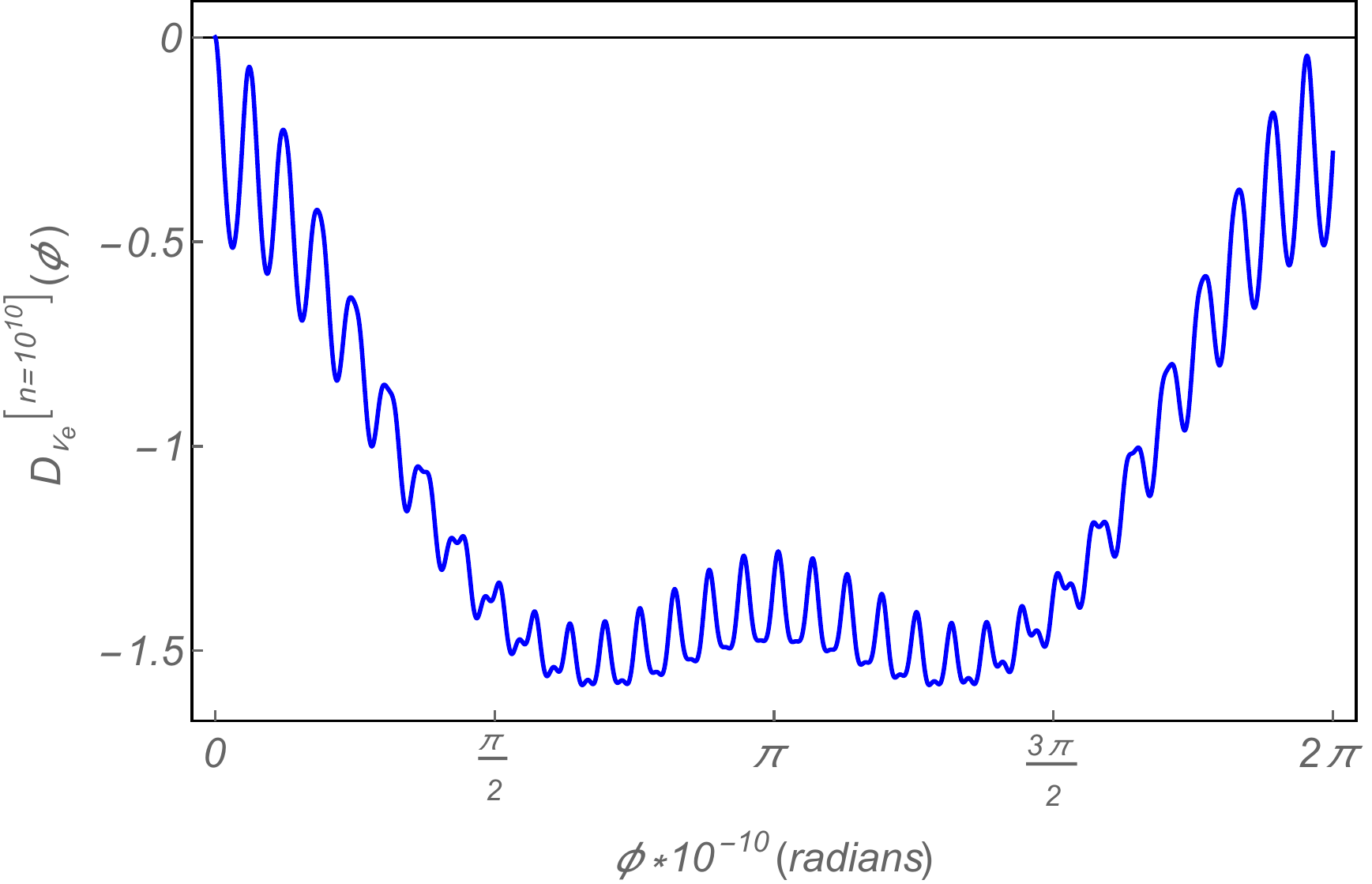}
	\caption{(Color online) Showing the the variation of the information dificit with the dimensionless parameter $\phi (=\frac{\Delta_{21}L}{2\hbar c E})$, for $n \approx 10^{10}$ in two (left) and three (right) flavor scenario of neutrino oscillations in vaccum. In three flavor case, the initial state is $\nu_e$. It is clear, that in the regime of large '$n$' and small '$\phi$', the value of the information deficit, $\mathcal{D}^{[n]}(\phi)$ is always negative.} \label{2F-3F-limit}
\end{figure*}

In conclusion, we have studied the entropic Leggett-Garg inequality for neutrinos in the context of neutrino oscillation and for B and K meson systems by using the formalism of open quantum systems. For the neutrino system, ELGI violation in both two and three flavor neutrino scenarios is studied. The strength of entropic violation (quantified by the information deficit $\mathcal{D}^{[n] }$) in three flavor case is roughly twice that in two flavor case. In two flavor case, the probabilities are independent of the initial state, so is the deficit parameter. In the three flavor case, the probabilities are initial state dependent, while the maximum negative value of $\mathcal{D}^{[n] }$ (measure of the extent of violation), shows variation with initial state dependence, only beyond the second decimal place. The extent of violation (characterized by the negative value of $\mathcal{D}^{[n] }$ ) increases with the increase in $n$-the number of observations/measurements made on the system. In the limit of  $n \rightarrow \infty$ and  $\phi \rightarrow 0$, as shown in Fig. (\ref{2F-3F-limit}), the value of information deficit $\mathcal{D}^{[n]}(\phi)$ is always negative, implying that  ELGI is always violated in this limit.\par
For the meson systems,  decoherence and CP violating effects are taken into account. We found that the ELGI is violated in $K$, $B_d$ and $B_s$ systems, such that the violation persists for a much longer time in K meson system as compared to the $B_d$ and $B_s$ systems. Enhancement in the violation with the increase in the number of measurements is found and is consistent with earlier works. The effect of decoherence is found to take the deficit parameter closer to its classical value zero.

\FloatBarrier

\end{document}